\documentstyle[seceq,preprint]{ptptex}

\newcommand{\bra}[1]{\langle {#1} |}     
\newcommand{\ket}[1]{| {#1} \rangle}     
\newcommand{\maru}[1]{\stackrel{\tiny\circ} {#1}} 

\title{
Deformed Boson Scheme including Conventional $q$-Deformation 
in Time-Dependent Variational Method. IV
}
\subtitle{The $su(2)_q$- and $su(1,1)_q$-Algebras in \\
Four Kinds 
of Boson Operators}    

\author{
Atsushi {\sc Kuriyama},${}^1$ 
Constan\c{c}a {\sc Provid\^encia},$^{2}$ \\
Jo\~ao da {\sc Provid\^encia},$^{2}$ Yasuhiko {\sc Tsue}$^{3}$ 
and Masatoshi {\sc Yamamura}${}^1$
}

\inst{
$^{1}$Faculty of Engineering, Kansai University, Suita 564-8680, Japan\\
$^{2}$Departamento de Fisica, Universidade de Coimbra, P-3000 Coimbra, 
Portugal\\
$^{3}$Physics Division, Faculty of Science, Kochi University, Kochi 780-8520, 
Japan
}

\recdate{
}

\abst{
Basic idea presented in Parts (I)$\sim$(III) for the deformed boson 
scheme is applied to the case of the $su(2)$- and $su(1,1)$-algebras 
for describing many-body systems consisting of four kinds of boson 
operators. A possible form of the coherent state given by the present 
authors is generalized and the $su(2)_q$- and the $su(1,1)_q$-algebras 
are obtained in the form expressed in terms of four kinds of boson operators. 
As an illustrative example of the application, the framework for 
describing thermal effects observed in two-level shell model under 
the pairing correlation is proposed. 
}

\begin{document}

\maketitle

\section{Introduction}

The present paper is a continuation of the last three papers (I), 
(II) and (III) for the deformed boson scheme.\cite{1} 
In Part (I), we focused our attention on a possible generalization of 
the conventional boson coherent state for the trial function in the 
time-dependent variational method. 
The scheme was formulated for the systems consisting of one kind 
of boson operator. The comparison of Part (I) with the paper by 
Penson and Solomon\cite{2} may be interesting. 
Part (II) was devoted to presenting the deformed boson scheme 
for the systems consisting of two kinds of boson operators. 
Through a certain procedure generalized from that in (I), 
we obtained the deformation of the $su(2)$- and $su(1,1)$-algebras 
given by Schwinger.\cite{3} 
Various forms of the $q$-deformed algebra were derived and, of course, 
one of the examples is based on the most popular deformation 
which is characterized by the function 
$[x]_q=(q^x-q^{-x})/(q-q^{-1})$. 
Furthermore, we obtained a transparent understanding for the treatment 
of a boson system interacting with an external harmonic oscillator 
in the framework of a certain pure state in the statistical sense.\cite{4} 
The idea of Ref.\citen{4} comes from the investigation on the 
$su(1,1)$-spin like behavior in the $su(2)$-spin system.\cite{5} 
Part (III) was aimed at formulating the $q$-deformation of the 
$su(2,1)$-algebra in the case where the generators 
are expressed in terms of the bilinear forms for three kinds of 
boson operators.\cite{6} 
With the help of this work, we were able to obtain a transparent understanding 
for the description of a boson system interacting with an external 
harmonic oscillator in terms of statistically mixed state.\cite{7} 
The above works\cite{4,5,6,7} are summarized in Ref.\citen{8}. 
In addition to the above three papers of this series about the deformed 
boson scheme, we published three short 
notes\cite{9,10} on some topics supplementing (I).

Under the above-mentioned situation, one of our next important tasks 
may be to formulate the $su(2)_q$- and the $su(1,1)_q$-algebras for 
the systems described in terms of four kinds of boson operators. 
There exist two reasons why we perform this task. 
As was already mentioned, in (II), we formulated the $su(2)_q$- 
and the $su(1,1)_q$-algebras in terms of two kinds of boson operators. 
Then, it may be interesting for formal theoretical aspects of the 
deformed boson scheme to investigate which parts of the framework 
given in (II) conserve their forms and which parts should be added 
newly to the forms in (II). The above is the first point. 
The second is related with physical problems. We know that the damped- 
and amplified-oscillation in one-dimensional harmonic oscillator 
can be described in terms of the $su(1,1)$-algebra in the Schwinger 
boson representation.\cite{3} In this case, the phase space doubling 
is essential and, two kinds of boson operators are necessary. 
The details can be found in Refs.\citen{11} and \citen{12}, 
together with Ref.\citen{8}. 
By extending the above idea, the present authors gave a possible method
to describe thermal and dissipative properties appearing 
in the $su(2)$-boson models consisting of one kind of boson operator.\cite{13} 
The typical example of the above $su(2)$-boson model is the shell model 
in which many nucleons move in single orbit under the pairing 
correlation. This model is treated in terms of the Holstein-Primakoff 
boson representation of the $su(2)$-algebra and the pairing rotation 
is the object of the investigation. 
Furthermore, on the basis of the viewpoint slightly different from 
that in Ref.\citen{13}, we made analysis for the 
$su(2)$-algebraic model such as that appearing in quantum optics.\cite{14} 
We showed that the deformed boson scheme given in Ref.\citen{9} 
reproduces the description quite transparently. The above-mentioned 
situation suggests us the following point : The deformed boson 
scheme may be helpful to the description of two-level 
shell model under the pairing correlation. 
In this model, not only the pairing rotation but also the pairing 
vibration is the object of the investigation and for the use of the 
Holstein-Primakoff boson representation, two kinds of boson operators are 
necessary. If we adopt the idea of the phase space doubling for the 
description of thermal properties, totally, four kinds of boson 
operators are necessary. The above is the second point.

The main aim of this paper is, as was already mentioned, to formulate 
the deformed boson scheme for treating the systems consisting of 
four kinds of boson operators. The fundamental element common to 
(I) $\sim$ (III) is to generalize the coherent state to the form 
appropriate for each case. In the exponential type, the coherent state 
is constructed in terms of some generators belonging to the algebra 
under consideration, which can be called the coherent state generating 
operators. Then, introducing certain function characterizing the 
deformation, we are able to give the generalized coherent state. 
With the use of these functions, the coherent state generating operators 
are deformed, and by calculating the commutation relations for 
these deformed operators, the deformed forms of the remaining generators 
are obtained. In Part (IV), i.e., this paper, this method is developed. 
In Ref.\citen{15}, the present authors investigated, in the frame 
of four kinds of boson operators, a possible form of the coherent state 
which is closely related to the $su(2)$- and the $su(1,1)$-algebras. 
This investigation is based on a generalized Schwinger representation 
containing the $su(M+1)$- and the $su(N,1)$-algebra.\cite{16} 
Since we are treating the case of four kinds of bosons, possible types of the 
deformation increase compared with the types shown in (II). 
One of the possible deformations gives us the Holstein-Primakoff 
boson representation of the $su(3)\times su(3)$-algebra in a certain 
symmetric representation and it is powerful for the understanding 
of thermal effects in two-level shell model under the pairing correlation. 
In this paper, the $su(2)$- and the $su(1,1)$-algebras appear in various 
notations, for example, $({\hat S}_{\pm}^0 , {\hat S}_0)$ and 
$({\hat T}_{\pm}^0 , {\hat T}_0)$. 
They obey the commutation relations 
\begin{subequations}\label{1-1}
\begin{eqnarray}
& &[ {\hat S}_{+}^0 , {\hat S}_-^0 ] = +2{\hat S}_0 \ , \qquad
[ {\hat S}_0 , {\hat S}_{\pm}^0 ]= \pm {\hat S}_{\pm}^0 \ , 
\label{1-1a}\\
& &[ {\hat T}_{+}^0 , {\hat T}_-^0 ] = -2{\hat T}_0 \ , \qquad
[ {\hat T}_0 , {\hat T}_{\pm}^0 ]= \pm {\hat T}_{\pm}^0 \ . 
\label{1-1b}
\end{eqnarray}
\end{subequations}
The Casimir operators are given as 
\begin{subequations}\label{1-2}
\begin{eqnarray}
& &{\hat {\mib S}}^2={\hat S}_0^2+({\hat S}_+^0{\hat S}_-^0
+{\hat S}_-^0{\hat S}_+^0)/2={\hat S}({\hat S}+1)\ , 
\label{1-2a}\\
& &{\hat {\mib T}}^2={\hat T}_0^2-({\hat T}_+^0{\hat T}_-^0
+{\hat T}_-^0{\hat T}_+^0)/2={\hat T}({\hat T}-1)\ . 
\label{1-2b}
\end{eqnarray}
\end{subequations}

The next section is devoted to recapitulating the $q$-deformation of the 
$su(2)$- and the $su(1,1)$-algebras in the case of two kinds of boson 
operators which was presented in (II). In \S 3, the $su(2)$- and 
the $su(1,1)$-algebras in the boson representation are recapitulated 
for the case of four kinds of bosons shown in Ref.\citen{15}. 
In \S 4, the deformation of the coherent state is discussed. 
The $su(2)_q$- and the $su(1,1)_q$-algebras are constructed in \S 5. 
In \S 6, an illustrative example of the application is investigated. 
Finally, in \S 7, as the concluding remarks, it is shown that 
the $q$-deformed algebra adopted in \S 6 is nothing but the Holstein-Primakoff 
representation of the $su(3)\times su(3)$-algebra in a certain symmetric 
representation. In Appendix, the explicit expression of the 
operator ${\hat T}$ defined in the relation (\ref{1-2b}) 
is given.

\section{The $su(2)$- and the $su(1,1)$-algebras for the case of the systems 
composed of two kinds of boson operators}

For convenience of later discussion, we first recapitulate the basic idea 
of the deformed boson scheme for the $su(2)$- and the $su(1,1)$-algebras 
for the case of two kinds of bosons. This scheme was developed in (II). 
In this section, we consider many-body systems composed of two kinds 
of boson operators $({\hat a} , {\hat a}^*)$ and $({\hat b} , {\hat b}^*)$. 
In this system, two sets of the operators $({\hat S}_{\pm}^0 , {\hat S}_0)$ 
and $({\hat T}_\pm^0, {\hat T}_0)$ are introduced in the form 
\begin{subequations}\label{2-1}
\begin{eqnarray}
& &{\hat S}_+^0={\hat a}^*{\hat b} \ , \qquad
{\hat S}_-^0={\hat b}^*{\hat a} \ , \qquad 
{\hat S}_0=(1/2)\cdot({\hat a}^*{\hat a}-{\hat b}^*{\hat b}) \ , 
\label{2-1a}\\
& &{\hat T}_+^0={\hat a}^*{\hat b}^* \ , \qquad
{\hat T}_-^0={\hat b}{\hat a} \ , \qquad 
{\hat T}_0=(1/2)\cdot({\hat a}^*{\hat a}+{\hat b}^*{\hat b}+1) \ . 
\label{2-1b}
\end{eqnarray}
\end{subequations}
The operators ${\hat S}$ and ${\hat T}$, which commute with 
$({\hat S}_{\pm}^0 , {\hat S}_0)$ and $({\hat T}_\pm^0, {\hat T}_0)$, 
respectively, are defined as 
\begin{subequations}\label{2-2}
\begin{eqnarray}
& &
{\hat S}=(1/2)\cdot({\hat b}^*{\hat b}+{\hat a}^*{\hat a}) \ , 
\label{2-2a}\\
& &
{\hat T}=(1/2)\cdot({\hat b}^*{\hat b}-{\hat a}^*{\hat a}+1) \ . 
\label{2-2b}
\end{eqnarray}
\end{subequations}
If we intend to investigate the $su(2)$-algebraic aspect of the systems, 
our main concern is the set $({\hat S}_\pm^0 , {\hat S}_0)$. 
Naturally, the case of the $su(1,1)$-algebraic aspect is concerned 
with the set $({\hat T}_\pm^0 , {\hat T}_0)$. 

The eigenstates of $({\hat S}+{\hat S}_0 , {\hat S})$ and 
$({\hat T}_0-{\hat T} , {\hat T})$ with the eigenvalues 
$(m_S , n/2)$ and $(m_T , n/2)$ are, respectively, given by 
\begin{subequations}\label{2-3}
\begin{eqnarray}
& &\ket{m_S ; n}=\sqrt{(n-m_S)!/n!}\left(\sqrt{m_S!}\right)^{-1}
\left({\hat S}_+^0\right)^{m_S}\ket{n} \ , 
\label{2-3a}\\
& &\ket{m_T ; n}=\sqrt{(n+m_T)!/n!}\left(\sqrt{m_T!}\right)^{-1}
\left({\hat T}_+^0\right)^{m_T}\ket{n} \ , 
\label{2-3b}
\end{eqnarray}
\end{subequations}
\begin{equation}\label{2-4}
\ket{n}=\left(\sqrt{n!}\right)^{-1}({\hat b}^*)^n \ket{0} \ . \qquad\quad
({\hat S}_-^0\ket{n}={\hat T}_-^0\ket{n}=0)
\end{equation}
If the eigenvalues of $({\hat S} , {\hat S}_0)$ and $({\hat T} , {\hat T}_0)$ 
are expressed as $(s , s_0)$ and $(t, t_0)$, respectively, the states 
(\ref{2-3a}) and (\ref{2-3b}) can be rewritten in the form 
\begin{subequations}\label{2-5}
\begin{eqnarray}
\ket{m_S ; n}&=&\ket{s,s_0}\nonumber\\
&=&\left(\sqrt{(s+s_0)!(s-s_0)!}\right)^{-1}
({\hat a}^*)^{s+s_0}({\hat b}^*)^{s-s_0}\ket{0} \ ,
\label{2-5a}\\
\ket{m_T;n}&=&\ket{t,t_0}\nonumber\\
&=&\left(\sqrt{(t_0-t)!(t_0+t-1)!}\right)^{-1}({\hat a}^*)^{t_0-t}
({\hat b}^*)^{t_0+t-1} \ket{0} \ , 
\label{2-5b}
\end{eqnarray}
\end{subequations}
\vspace{-0.4cm}
\begin{subequations}\label{2-6}
\begin{eqnarray}
& &m_S=s+s_0 \ , \qquad  n=2s \ , 
\label{2-6a}\\
& &m_T=t_0-t \ , \qquad n=2t \ .
\label{2-6b}
\end{eqnarray}
\end{subequations}
It should be noted that, in the state (\ref{2-3b}), the eigenvalue of 
${\hat T}$ is positive. Under the superposition of the states 
(\ref{2-3a}) and (\ref{2-3b}), we are able to define the following 
wave packets : 
\begin{subequations}\label{2-7}
\begin{eqnarray}
& &\ket{c_S^0}=\left(\sqrt{\Gamma_S^0}\right)^{-1}
\exp\left(\gamma_S {\hat S}_+^0\right)\ket{\delta} \ , \qquad\qquad
\label{2-7a}\\
& &\ket{c_T^0}=\left(\sqrt{\Gamma_T^0}\right)^{-1}
\exp\left(\gamma_T {\hat T}_+^0\right)\ket{\delta} \ , 
\label{2-7b}
\end{eqnarray}
\end{subequations}
\begin{equation}\label{2-8}
\ket{\delta}=\exp\left(\delta{\hat b}^*\right)\ket{0} \ .
\qquad ({\hat S}_-^0\ket{\delta}={\hat T}_-^0\ket{\delta}=0)
\end{equation}
Here, $\gamma_S$, $\gamma_T$ and $\delta$ denote complex parameters 
and $\Gamma_S^0$ and $\Gamma_T^0$ are the normalization 
constants given by 
\begin{subequations}\label{2-9}
\begin{eqnarray}
& &\Gamma_S^0=\exp \left((1+|\gamma_S|^2)|\delta|^2\right) \ , 
\label{2-9a}\\
& &\Gamma_T^0=(1-|\gamma_T|^2)^{-1}
\exp \left((1-|\gamma_T|^2)^{-1}|\delta|^2\right) \ . 
\label{2-9b}
\end{eqnarray}
\end{subequations}
In (II), we introduced the following operators : 
\begin{subequations}\label{2-10}
\begin{eqnarray}
& &{\hat \gamma}_S^0={\hat S}_-^0({\hat N}_b+1+\epsilon)^{-1} \ , 
\qquad {\hat \delta}_S^0={\hat b} \ , 
\label{2-10a}\\
& &{\hat \gamma}_T^0=({\hat N}_b+1+\epsilon)^{-1}{\hat T}_-^0 \ , 
\qquad {\hat \delta}_T^0=[1-{\hat N}_a({\hat N}_b+1+\epsilon)^{-1}]
{\hat b} \ . 
\label{2-10b}
\end{eqnarray}
\end{subequations}
Here, $\epsilon$ denotes an infinitesimal parameter. The operators 
${\hat N}_a$ and ${\hat N}_b$ mean the boson numbers and they are 
related with the generators in the following form : 
\begin{equation}\label{2-11}
\ \ {\hat N}_a={\hat a}^*{\hat a} \ , \qquad 
{\hat N}_b={\hat b}^*{\hat b} \ , \qquad\qquad\qquad
\end{equation}
\vspace{-1.1cm}
\begin{subequations}\label{2-12}
\begin{eqnarray}
& &{\hat N}_a={\hat S}+{\hat S}_0 \ , \qquad {\hat N}_b={\hat S}-{\hat S}_0 \ ,
\label{2-12a}\\
& &{\hat N}_a={\hat T}_0-{\hat T} \ , 
\qquad {\hat N}_b={\hat T}_0+{\hat T}-1 \ .
\label{2-12b}
\end{eqnarray}
\end{subequations}
Then, the form (\ref{2-10}) can be rewritten as 
\begin{subequations}\label{2-13}
\begin{eqnarray}
& &{\hat \gamma}_S^0=({\hat S}-{\hat S}_0+\epsilon)^{-1}{\hat S}_-^0 \ , 
\qquad {\hat \delta}_S^0={\hat b} \ , 
\label{2-13a}\\
& &{\hat \gamma}_T^0=({\hat T}_0+{\hat T}+\epsilon)^{-1}{\hat T}_-^0 \ , 
\qquad {\hat \delta}_T^0={\hat b}-{\hat a}^*
({\hat T}_0+{\hat T}+\epsilon)^{-1}{\hat T}_-^0 \ . 
\label{2-13b}
\end{eqnarray}
\end{subequations}
As was shown in (II), we have the relations 
\begin{subequations}\label{2-14}
\begin{eqnarray}
& &{\hat \gamma}_S^0\ket{c_S^0}
=\gamma_S[1-\epsilon({S}-{S}_0+\epsilon)^{-1}]\ket{c_S^0}\ , 
\quad 
{\hat \delta}_S^0\ket{c_S^0}=\delta\ket{c_S^0} \ , \qquad\quad
\label{2-14a}\\
& &{\hat \gamma}_T^0\ket{c_T^0}
=\gamma_T\ket{c_T^0}\ , 
\qquad\qquad\qquad\qquad\qquad 
{\hat \delta}_T^0\ket{c_T^0}=\delta\ket{c_T^0} \ . 
\label{2-14b}
\end{eqnarray}
\end{subequations}
The above relations tell us that the states $\ket{c_S^0}$ and $\ket{c_T^0}$ 
can be regarded as the $su(2)$- and the $su(1,1)$-spin coherent states, 
respectively.

As a possible deformation of $\ket{c_S^0}$ and $\ket{c_T^0}$, in (II), 
we investigated the following form : 
\begin{subequations}\label{2-15}
\begin{eqnarray}
& &\ket{c_S}=\sqrt{\Gamma_S^0/\Gamma_S}\cdot f_S({\hat N}_a)
g_S({\hat N}_b)^{-1}\ket{c_S^0} \ , 
\label{2-15a}\\
& &\ket{c_T}=\sqrt{\Gamma_T^0/\Gamma_T}\cdot f_T({\hat N}_a)
g_T({\hat N}_b)\ket{c_T^0} \ . 
\label{2-15b}
\end{eqnarray}
\end{subequations}
Here, $\Gamma_S$ and $\Gamma_T$ denote the normalization constants. 
The deformation of $\ket{c_S^0}$ and $\ket{c_T^0}$ are characterized 
by the functions $f_S$, $g_S$, $f_T$ and $g_T$ and the details of these 
functions were mentioned in (II). With the use of the relations 
(\ref{2-12a}) and (\ref{2-12b}), $\ket{c_S}$ and $\ket{c_T}$ can be 
expressed as 
\begin{subequations}\label{2-16}
\begin{eqnarray}
& &\ket{c_S}=\sqrt{\Gamma_S^0/\Gamma_S}\cdot \Omega_S({\hat S}, {\hat S}_0)
\ket{c_S^0} \ , \nonumber\\
& &\ \ \Omega_S({\hat S},{\hat S}_0)=f_S({\hat S}+{\hat S}_0)
g_S({\hat S}-{\hat S}_0)^{-1} \ , 
\label{2-16a}\\
& &\ket{c_T}=\sqrt{\Gamma_T^0/\Gamma_T}\cdot \Omega_T({\hat T}, {\hat T}_0)
\ket{c_T^0} \ , \nonumber\\
& &\ \ \Omega_T({\hat T},{\hat T}_0)=f_T({\hat T}_0-{\hat T})
g_T({\hat T}_0+{\hat T}-1) \ . 
\label{2-16b}
\end{eqnarray}
\end{subequations}
In the present case, ${\hat \gamma}_S$, ${\hat \delta}_S$, ${\hat \gamma}_T$ 
and ${\hat \delta}_T$ can be defined in the form 
\begin{subequations}\label{2-17}
\begin{eqnarray}
& &{\hat \gamma}_S=\Omega_S({\hat S},{\hat S}_0){\hat \gamma}_S^0
\Omega_S({\hat S},{\hat S}_0)^{-1} \ , \qquad
{\hat \delta}_S=\Omega_S({\hat S},{\hat S}_0){\hat \delta}_S^0
\Omega_S({\hat S},{\hat S}_0)^{-1} \ ,
\label{2-17a}\\
& &{\hat \gamma}_T=\Omega_T({\hat T},{\hat T}_0){\hat \gamma}_T^0
\Omega_T({\hat T},{\hat T}_0)^{-1} \ , \qquad
{\hat \delta}_T=\Omega_T({\hat T},{\hat T}_0){\hat \delta}_T^0
\Omega_T({\hat T},{\hat T}_0)^{-1} \ .\qquad\quad
\label{2-17b}
\end{eqnarray}
\end{subequations}
The operation of ${\hat \gamma}_S$, ${\hat \delta}_S$, ${\hat \gamma}_T$ 
and ${\hat \delta}_T$ on the states $\ket{c_S}$ and $\ket{c_T}$ gives us 
the same forms as those shown in the relations (\ref{2-14a}) and 
(\ref{2-14b}), respectively. 
In this sense, $\ket{c_S}$ and $\ket{c_T}$ are still coherent states 
for these operators.

With the use of the functions $f_S$, $g_S$, $f_T$ and $g_T$, we are 
able to obtain the $q$-deformation of the $su(2)$- and the 
$su(1,1)$-algebras. First, we define ${\hat S}_\pm$ and ${\hat T}_\pm$ 
in the form 
\begin{subequations}\label{2-18}
\begin{eqnarray}
& &{\hat S}_+=\Omega_S({\hat S},{\hat S}_0)^{-1}
{\hat S}_+^0 \Omega_S({\hat S}, {\hat S}_0) \ , \qquad
{\hat S}_-=\Omega_S({\hat S},{\hat S}_0)
{\hat S}_-^0 \Omega_S({\hat S}, {\hat S}_0)^{-1} \ , 
\label{2-18a}\\
& &{\hat T}_+=\Omega_T({\hat T},{\hat T}_0)^{-1}
{\hat T}_+^0 \Omega_T({\hat T}, {\hat T}_0) \ , \qquad
{\hat T}_-=\Omega_T({\hat T},{\hat T}_0)
{\hat T}_-^0 \Omega_T({\hat T}, {\hat T}_0)^{-1} \ . \qquad\quad
\label{2-18b}
\end{eqnarray}
\end{subequations}
Then, $[2{\hat S}_0]_q$ and $[2{\hat T}_0]_q$ are given as follows : 
\begin{subequations}\label{2-19}
\begin{eqnarray}
[2{\hat S}_0]_q
&=&[{\hat S}_+ , {\hat S}_-] \nonumber\\
&=&[{\hat S}+{\hat S}_0]_{f_S}[{\hat S}-{\hat S}_0+1]_{g_S}
-[{\hat S}+{\hat S}_0+1]_{f_S}[{\hat S}-{\hat S}_0]_{g_S} \ , 
\label{2-19a}\\
{[}2{\hat T}_0]_q
&=&[{\hat T}_- , {\hat T}_+] \nonumber\\
&=&[{\hat T}_0-{\hat T}+1]_{f_T}[{\hat T}_0+{\hat T}]_{g_T}
-[{\hat T}_0-{\hat T}]_{f_T}[{\hat T}_0+{\hat T}-1]_{g_T} \ . 
\label{2-19b}
\end{eqnarray}
\end{subequations}
Here, $[x]_{f_S}$, $[x]_{g_S}$, $[x]_{f_T}$ and $[x]_{g_T}$ are 
defined as 
\begin{subequations}\label{2-20}
\begin{eqnarray}
& &[x]_{f_S}=x f_S(x)^{-2}f_S(x-1)^2 \ , \qquad
[x]_{g_S}=x g_S(x)^{-2}g_S(x-1)^2 \ , 
\label{2-20a}\\
& &[x]_{f_T}=x f_T(x)^{-2}f_T(x-1)^2 \ , \qquad
[x]_{g_T}=x g_T(x)^{-2}g_T(x-1)^2 \ . 
\label{2-20b}
\end{eqnarray}
\end{subequations}
The details of the above treatment are shown in (II). 
Furthermore, the case in which the functions $f(x)$ and $g(x)$ have 
singularities was discussed in Ref.\citen{9}. 
In this case, the use of a certain projection operator is essential.

\section{The $su(2)$- and the $su(1,1)$-algebras in the boson representation 
for four kinds of boson operators}

As was mentioned in \S 1, the aim of this paper is to apply the basic 
idea of the deformed boson scheme presented in the previous section 
to many-body systems consisting of four kinds of boson operators. 
For such boson systems, we can imagine various algebras. For example, 
the $su(4)$-algebra in a symmetric boson representation may be the most 
popular one. In the present paper, we investigate the $su(2)$- and 
the $su(1,1)$-algebras. For this aim, the formalism developed by the 
present authors in Ref.\citen{15} is useful. From the above reason, 
we, first, recapitulate the main part of Ref.\citen{15} in a form 
slightly different from the original one with the new notations. 

We denote the four kinds of boson operators as 
$({\hat a}_\pm, {\hat a}_{\pm}^*)$ and $({\hat b}_\pm, {\hat b}_\pm^*)$. 
With the use of these operators, $({\hat S}_\pm^0 , {\hat S}_0)$ and 
$({\hat T}_\pm^0 , {\hat T}_0)$ can be expressed as 
\begin{subequations}\label{3-1}
\begin{eqnarray}
& &{\hat S}_+^0={\hat a}_+^*{\hat b}_+ + {\hat a}_-^*{\hat b}_- \ , 
\qquad
{\hat S}_-^0={\hat b}_+^*{\hat a}_+ + {\hat b}_-^*{\hat a}_- \ , \nonumber\\
& &{\hat S}_0=(1/2)\cdot ({\hat a}_+^*{\hat a}_+
-{\hat b}_+^*{\hat b}_+)+(1/2)\cdot ({\hat a}_-^*{\hat a}_-
-{\hat b}_-^*{\hat b}_-) \ , 
\label{3-1a}\\
& &{\hat T}_+^0={\hat a}_+^*{\hat b}_-^* - {\hat a}_-^*{\hat b}_+^* \ , 
\qquad
{\hat T}_-^0={\hat b}_-{\hat a}_+ - {\hat b}_+{\hat a}_- \ , \nonumber\\
& &{\hat T}_0=(1/2)\cdot ({\hat a}_+^*{\hat a}_+
+{\hat b}_-^*{\hat b}_-+1)+(1/2)\cdot ({\hat a}_-^*{\hat a}_-
+{\hat b}_+^*{\hat b}_++1) \ . 
\label{3-1b}
\end{eqnarray}
Further, we introduce new $su(2)$-generators $({\hat R}_\pm^0 , {\hat R}_0)$ 
which play an auxiliary but essential role in the present formalism : 
\begin{eqnarray}\label{3-1c}
& &{\hat R}_+^0={\hat a}_+^*{\hat a}_- + {\hat b}_+^*{\hat b}_- \ , 
\qquad
{\hat R}_-^0={\hat a}_-^*{\hat a}_+ + {\hat b}_-^*{\hat b}_+ \ , \nonumber\\
& &{\hat R}_0=(1/2)\cdot({\hat a}_+^*{\hat a}_+ - {\hat a}_-^*{\hat a}_-)
+(1/2)\cdot({\hat b}_+^*{\hat b}_+ - {\hat b}_-^*{\hat b}_-) \ .
\end{eqnarray}
\end{subequations}
It should be noted that we have the relation 
\begin{eqnarray}\label{3-2}
& &[\ \hbox{\rm any\ of}\ ({\hat R}_\pm^0 , {\hat R}_0) \ , \ 
\hbox{\rm any\ of}\ ({\hat S}_\pm^0 , {\hat S}_0) \ ]=0 \ , \nonumber\\
& &[\ \hbox{\rm any\ of}\ ({\hat R}_\pm^0 , {\hat R}_0) \ , \ 
\hbox{\rm any\ of}\ ({\hat T}_\pm^0 , {\hat T}_0) \ ]=0 \ , \nonumber\\
& &[\ \hbox{\rm any\ of}\ ({\hat S}_\pm^0 , {\hat S}_0) \ , \ 
\hbox{\rm any\ of}\ ({\hat T}_\pm^0 , {\hat T}_0) \ ]=0 \ . 
\end{eqnarray}
The Casimir operators ${\hat {\mib R}}^2(={\hat R}({\hat R}+1))$, 
${\hat {\mib S}}^2(={\hat S}({\hat S}+1))$ and 
${\hat {\mib T}}^2(={\hat T}({\hat T}-1))$ are identical to each other 
and, then, as a possible choice, we can set up the relation 
\begin{equation}\label{3-3}
{\hat R}={\hat S}={\hat T}-1 \ .
\end{equation}
Since the eigenvalues of ${\hat R}$ and ${\hat S}$ are always positive, 
the eigenvalue of ${\hat T}$ is always larger than 1. 
In Appendix, the explicit form of ${\hat T}$ is given in terms 
of $({\hat T}_\pm^0,{\hat T}_0)$, i.e., $({\hat a}_\pm, {\hat a}_\pm^*)$ 
and $({\hat b}_\pm, {\hat b}_\pm^*)$.

For the above system, we consider a set of the states 
$\{\ket{m_R\ m_S\ m_T ; n}\}$ : 
\begin{eqnarray}
& &\ket{m_R\ m_S\ m_T ; n}
=\left(\sqrt{N_{m_Rm_Sm_T;n}}\right)^{-1}\nonumber\\
& &\qquad\qquad
\times (\sqrt{m_R!})^{-1}(\sqrt{m_S!})^{-1}
(\sqrt{m_T!})^{-1}({\hat R}_+^0)^{m_R}({\hat S}_+^0)^{m_S}
({\hat T}_+^0)^{m_T}\ket{n} \ , 
\label{3-4}\\
& &\ket{n}=(\sqrt{n!})^{-1}({\hat b}_-^*)^n \ket{0} \ . \qquad
({\hat R}_-^0\ket{n}={\hat S}_-^0\ket{n}={\hat T}_-^0\ket{n}=0)
\end{eqnarray}
Here, $N_{m_Rm_Sm_T;n}$ denotes the normalization constant : 
\begin{equation}\label{3-6}
N_{m_Rm_Sm_T;n}=[n!/(n-m_R)!]\cdot[n!/(n-m_S)!]\cdot[(n+1+m_T)!/(n+1)!] \ .
\end{equation}
The state (\ref{3-4}) is a simultaneous eigenstate of the operators 
${\hat R}+{\hat R}_0$, ${\hat S}+{\hat S}_0$, ${\hat T}_0-{\hat T}$ and 
${\hat R}={\hat S}={\hat T}-1$ which are given as 
\begin{equation}\label{3-7}
\hbox{\rm the\ eigenvalue\ of}\ 
\cases{{\hat R}+{\hat R}_0 \quad \hbox{\rm is}\quad m_R \ , \cr
{\hat S}+{\hat S}_0 \quad \ \hbox{\rm is}\quad m_S \ , \cr
{\hat T}_0-{\hat T} \quad \ \hbox{\rm is}\quad m_T \ , \cr
{\hat R}={\hat S}={\hat T}-1 \quad \hbox{\rm is}\quad n/2 \ .}
\end{equation}
Therefore, $\{\ket{m_R\ m_S\ m_T ;n}\}$ forms an orthogonal and complete 
set. Under the superposition of the states (\ref{3-4}), we define 
the following wave packet : 
\begin{eqnarray}
& &\ket{c^0}=\left(\sqrt{\Gamma^0}\right)^{-1}
\exp\left(\gamma_R{\hat R}_+^0\right)
\exp\left(\gamma_S{\hat S}_+^0\right)
\exp\left(\gamma_T{\hat T}_+^0\right)\ket{\delta} \ , 
\label{3-8}\\
& &\ket{\delta}=\exp(\delta{\hat b}_-^*)\ket{0}\ . \qquad
({\hat R}_-^0\ket{\delta}={\hat S}_-^0\ket{\delta}={\hat T}_-^0\ket{\delta}
=0)
\label{3-9}
\end{eqnarray}
Here, $\gamma_R$, $\gamma_S$, $\gamma_T$ and $\delta$ denote complex 
parameters and $\Gamma^0$ is the normalization constant given as 
\begin{equation}\label{3-10}
\Gamma^0=(1-|\gamma_T|^2)\exp\left(
(1+|\gamma_R|^2)(1+|\gamma_S|^2)(1-|\gamma_T|^2)^{-1}|\delta|^2\right) \ . 
\end{equation}
In the same manner as that in \S 2, we can introduce the following 
operators : 
\begin{eqnarray}\label{3-11}
& &{\hat \gamma}_R^0=({\hat R}-{\hat R}_0+\epsilon)^{-1}{\hat R}_-^0 \ ,
\nonumber\\
& &{\hat \gamma}_S^0=({\hat S}-{\hat S}_0+\epsilon)^{-1}{\hat S}_-^0 \ ,
\nonumber\\
& &{\hat \gamma}_T^0=({\hat T}_0+{\hat T}+\epsilon)^{-1}{\hat T}_-^0 \ ,
\nonumber\\
& &{\hat \delta}^0={\hat b}_- - 
{\hat a}_+^*({\hat T}_0+{\hat T}+\epsilon)^{-1}{\hat T}_-^0 \ .
\end{eqnarray}
They satisfy the relations
\begin{eqnarray}\label{3-12}
& &{\hat \gamma}_R^0\ket{c^0}
=\gamma_R[1-\epsilon({R}-{R}_0+\epsilon)^{-1}]\ket{c^0} \ ,
\nonumber\\
& &{\hat \gamma}_S^0\ket{c^0}
=\gamma_S[1-\epsilon({S}-{S}_0+\epsilon)^{-1}]\ket{c^0} \ ,
\nonumber\\
& &{\hat \gamma}_T^0\ket{c^0}
=\gamma_T\ket{c^0} \ ,
\nonumber\\
& &{\hat \delta}^0\ket{c^0}=\delta\ket{c^0} \ .
\end{eqnarray}
The relations (\ref{3-12}) tell us that the state $\ket{c^0}$ can be 
regarded as a coherent state for ${\hat \gamma}_R^0$, ${\hat \gamma}_S^0$, 
${\hat \gamma}_T^0$ and ${\hat \delta}^0$. It should be noted that 
${\hat R}$, ${\hat S}$ and ${\hat T}$ in the definition (\ref{3-11}) 
can be expressed in terms of $({\hat a}_\pm, {\hat a}_\pm^*)$ 
and $({\hat b}_\pm, {\hat b}_\pm^*)$. 
The operators ${\hat R}_+^0$, ${\hat S}_+^0$ and ${\hat T}_+^0$ 
can be regarded as operators generating the coherent state.

\section{Deformation of the coherent state $\ket{c^0}$}

We are now in a stage to give a possible deformation of the system 
which was discussed in \S 3. First, we note the following point : 
As can be seen in the forms (\ref{3-1a}) and (\ref{3-1b}), 
$({\hat S}_\pm^0 , {\hat S}_0)$ and $({\hat T}_\pm^0 , {\hat T}_0)$ 
consist of two $su(2)$- and the $su(1,1)$-spins, 
$({\hat I}_\pm^0+{\hat J}_\pm^0 , {\hat I}_0+{\hat J}_0)$ and 
$({\hat K}_\pm^0-{\hat L}_\pm^0 , {\hat K}_0+{\hat L}_0)$, respectively. 
The four sets $({\hat I}_\pm^0 , {\hat I}_0)$, 
$({\hat J}_\pm^0 , {\hat J}_0)$, $({\hat K}_\pm^0 , {\hat K}_0)$ and 
$({\hat L}_\pm^0 , {\hat L}_0)$ can be expressed in terms of 
$({\hat a}_+ , {\hat a}_+^* ; {\hat b}_+ , {\hat b}_+^*)$, 
$({\hat a}_- , {\hat a}_-^* ; {\hat b}_- , {\hat b}_-^*)$, 
$({\hat a}_+ , {\hat a}_+^* ; {\hat b}_- , {\hat b}_-^*)$ and 
$({\hat a}_- , {\hat a}_-^* ; {\hat b}_+ , {\hat b}_+^*)$, respectively. 
Each is nothing but the Schwinger representation shown in \S 2. 
Therefore, following the relations (\ref{2-2a}) and (\ref{2-2b}), 
we are able to define the operators ${\hat I}$, ${\hat J}$, ${\hat K}$ 
and ${\hat L}$ in the form 
\begin{subequations}\label{4-1}
\begin{eqnarray}
& &{\hat I}=(1/2)\cdot({\hat b}_+^*{\hat b}_+ + {\hat a}_+^*{\hat a}_+) \ , 
\qquad\quad
{\hat J}=(1/2)\cdot({\hat b}_-^*{\hat b}_- + {\hat a}_-^*{\hat a}_-) \ , 
\label{4-1a}\\
& &{\hat K}=(1/2)\cdot({\hat b}_-^*{\hat b}_- - 
{\hat a}_+^*{\hat a}_+ + 1) \ , 
\quad
{\hat L}=(1/2)\cdot({\hat b}_+^*{\hat b}_+ - 
{\hat a}_-^*{\hat a}_- + 1) \ . 
\label{4-1b}
\end{eqnarray}
\end{subequations}
The properties of the above operators are self-evident. 
With the use of the above operators, we can set up the eigenvalue equations : 
\begin{subequations}\label{4-2}
\begin{eqnarray}
& &{\hat I}\ket{i,i_0}_I=i\ket{i,i_0}_I \ , \qquad\qquad
{\hat I}_0\ket{i,i_0}_I=i_0\ket{i,i_0}_I \ , \nonumber\\
& &{\hat J}\ket{j,j_0}_J=j\ket{j,j_0}_J \ , \qquad\quad
{\hat J}_0\ket{j,j_0}_J=j_0\ket{j,j_0}_J \ , 
\label{4-2a}\\
& &{\hat K}\ket{k,k_0}_K=k\ket{k,k_0}_K \ , \qquad
{\hat K}_0\ket{k,k_0}_K=k_0\ket{k,k_0}_K \ , \nonumber\\
& &{\hat L}\ket{l,l_0}_L=l\ket{l,l_0}_L \ , \qquad\qquad
{\hat L}_0\ket{l,l_0}_L=l_0\ket{l,l_0}_L \ . 
\label{4-2b}
\end{eqnarray}
\end{subequations}
In \S 2, we restricted ourselves to the positive $t$. However, 
in this section, $k$ and $l$ are not restricted to the positive values. 
Since $\ket{m_R\ m_S\ m_T ; n}$ is also the eigenstate of 
$({\hat I}, {\hat J}, {\hat S}, {\hat S}_0)$ and $({\hat K}, {\hat L}, 
{\hat T}, {\hat T}_0)$, we have the following relations : 
\begin{subequations}\label{4-3}
\begin{eqnarray}
& &\ket{m_R\ m_S\ m_T ; n}=\ket{ij;ss_0}_S
=\sum_{i_0, j_0}\bra{ii_0jj_0}ss_0\rangle_S 
\ket{i,i_0}_I\times\ket{j,j_0}_J \ , 
\label{4-3a}\\
& &\ket{m_R\ m_S\ m_T ; n}=\ket{kl;tt_0}_T
=\sum_{k_0, l_0}\bra{kk_0ll_0}tt_0\rangle_T 
\ket{k,k_0}_K\times\ket{l,l_0}_L \ , \qquad\quad
\label{4-3b}
\end{eqnarray}
\end{subequations}
\vspace{-0.8cm}
\begin{subequations}\label{4-4}
\begin{eqnarray}
& &m_R=i-j+s \ , \qquad m_T=i+j-s \ , \qquad m_S=s+s_0 \ , \qquad n=2s \ , 
\label{4-4a}\\
& &m_R=t-k+l-1 \ , \quad m_S=t-k-l \ , \quad m_T=t_0-t \ , \quad n=2t-2 \ . 
\quad 
\label{4-4b}
\end{eqnarray}
\end{subequations}
Here, the quantities $\bra{ii_0jj_0}ss_0\rangle_S$ and 
$\bra{kk_0ll_0}tt_0\rangle_T$ denote the $su(2)$- and the 
$su(1,1)$-Clebsh-Gordan coefficients, respectively. The relations 
(\ref{4-4a}) and (\ref{4-4b}) tell us that the state (\ref{3-4}) 
gives us the coupling rule of the $su(2)$- and the $su(1,1)$-spins : 
\begin{subequations}\label{4-5}
\begin{eqnarray}
& &s=s_m,\ s_m+1, \ s_m+2, \cdots, (i+j) \ , \qquad s_m=|i-j| \ , \nonumber\\
& &s_0=-s,\ -s+1, \cdots, s-1,\ s \ , 
\label{4-5a}\\
& &t=t_m,\ t_m+1,\ t_m+2, \cdots \ , 
\quad t_m=|k-1/2|+|l-1/2|+1 \ , \nonumber\\
& &t_0=t,\ t+1,\ t+2 , \cdots \ . 
\label{4-5b}
\end{eqnarray}
\end{subequations}
The details can be seen in Ref.\citen{15}.

Under the above preparation, let us investigate the deformation of 
the coherent state $\ket{c^0}$ shown in the relation (\ref{3-8}) 
and (\ref{3-9}). By expanding $\ket{c^0}$ in terms of the orthogonal 
set $\{\ket{m_R\ m_S\ m_T ; n}\}$ in the relation (\ref{3-4}), we have 
the following form : 
\begin{eqnarray}
& &\ket{c^0}=\sum_{m_R,m_S,m_T,n}C_{m_R m_S m_T n}\ket{m_R\ m_S\ m_T; n} \ ,
\label{4-6}\\
& &C_{m_R m_S m_T n}=\left(\sqrt{\Gamma^0}\right)^{-1}
(\sqrt{m_R!m_S!m_T!n!})^{-1}\sqrt{N_{m_Rm_Sm_T;n}} \nonumber\\
& &\qquad\qquad\qquad\times
(\gamma_R)^{m_R}(\gamma_S)^{m_S}(\gamma_T)^{m_T}(\delta)^n \ .
\label{4-7}
\end{eqnarray}
With the use of the relations (\ref{4-3a}), (\ref{4-3b}), (\ref{4-4a}) 
and (\ref{4-4b}), we have two forms, which are formally different from 
each other, for $\ket{c^0}$. We denote them 
as $\ket{c_S^0}$ and $\ket{c_T^0}$ : 
\begin{subequations}\label{4-8}
\begin{eqnarray}
& &\ket{c_S^0}=\sum_{i,j,s,s_0}C_{ij;ss_0}(S)\ket{ij;ss_0}_S \ , 
\label{4-8a}\\
& &\ket{c_T^0}=\sum_{k,l,t,t_0}C_{kl;tt_0}(T)\ket{kl;tt_0}_T \ . 
\label{4-8b}
\end{eqnarray}
\end{subequations}
The above two states are essentially the same as each other. However, 
in the case where we consider the deformation, the formal difference becomes 
essential. The coefficients $C_{ij;ss_0}(S)$ and $C_{kl;tt_0}(T)$ 
denote the amplitudes of the state with the $su(2)$-spin $(s,s_0)$ 
coupled with two $su(2)$-spins $(i,i_0)$ and $(j,j_0)$ and of the state 
with the $su(1,1)$-spin $(t,t_0)$ coupled with the $su(1,1)$-spins 
$(k,k_0)$ and $(l, l_0)$, respectively. 
Therefore, if we are interested in the $su(2)$-algebraic aspect 
in the present system, we should perform the deformation of $\ket{c^0}$ 
based on the form (\ref{4-8a}). On the other hand, if in the case of 
the $su(1,1)$-algebraic aspect, we should perform the deformation 
based on the form (\ref{4-8b}). The two amplitudes $C_{ij;ss_0}(S)$ 
and $C_{kl;tt_0}(T)$ depend on the $(ij;ss_0)$ and $(kl;tt_0)$, 
respectively. Then, borrowing the idea in the case developed in \S 2, 
we make the following deformation : 
\begin{subequations}\label{4-9}
\begin{eqnarray}
& &\ket{c_S}=\left(\sqrt{\Gamma_S}\right)^{-1}
\sum_{i,j,s,s_0}C_{ij;ss_0}(S)\Omega_S(ij;ss_0)\ket{ij;ss_0}_S \ , 
\label{4-9a}\\
& &\ket{c_T}=\left(\sqrt{\Gamma_T}\right)^{-1}
\sum_{k,l,t,t_0}C_{kl;tt_0}(T)\Omega_T(kl;tt_0)\ket{kl;tt_0}_T \ . 
\label{4-9b}
\end{eqnarray}
\end{subequations}
Here, $\Gamma_S$ and $\Gamma_T$ denote the normalization constants. 
The functions $\Omega_S(ij;ss_0)$ and $\Omega_T(kl;tt_0)$ are defined 
as follows : 
\begin{subequations}\label{4-10}
\begin{eqnarray}
& &\Omega_S(ij;ss_0)=d_S(i)e_S(j)f_S(s+s_0)g_S(s-s_0)^{-1} \ , 
\label{4-10a}\\
& &\Omega_T(kl;tt_0)=d_T(k-1/2)e_T(l-1/2)f_T(t_0-t)g_T(t_0+t-2) \ . 
\label{4-10b}
\end{eqnarray}
\end{subequations}
The quantities $i$ and $j$ take the values $0, 1/2, 1, 3/2, \cdots$. 
However, in the case of $(k,l)$, the negative integers and half-integers 
are also permitted. The present deformation is characterized by the sets 
of the functions $(d_S, e_S, f_S, g_S)$ and $(d_T, e_T, f_T, g_T)$. 
We can see that the amplitudes change from $C_{ij;ss_0}(S)$ and 
$C_{kl;tt_0}(T)$ to $C_{ij;ss_0}(S)\Omega_S(ij;ss_0)$ and 
$C_{kl;tt_0}(T)\Omega_T(kl;tt_0)$, respectively. 
The states $\ket{c_S}$ and $\ket{c_T}$ can be rewritten as 
\begin{subequations}\label{4-11}
\begin{eqnarray}
& &\ket{c_S}=\sqrt{\Gamma^0/\Gamma_S}\ {\hat \Omega}_S\ket{c_S^0} \ ,
\label{4-11a}\\
& &\ket{c_T}=\sqrt{\Gamma^0/\Gamma_T}\ {\hat \Omega}_T\ket{c_T^0} \ ,
\qquad\qquad\qquad\qquad\qquad\qquad\qquad
\label{4-11b}
\end{eqnarray}
\end{subequations}
\vspace{-0.5cm}
\begin{subequations}\label{4-12}
\begin{eqnarray}
{\hat \Omega}_S&=&\Omega_S({\hat I}{\hat J};{\hat S}{\hat S}_0) \nonumber\\
&=&d_S({\hat I})e_S({\hat J})f_S({\hat S}+{\hat S}_0)
g_S({\hat S}-{\hat S}_0)^{-1} \ , 
\label{4-12a}\\
{\hat \Omega}_T&=&\Omega_T({\hat K}{\hat L};{\hat T}{\hat T}_0)\nonumber\\
&=&d_T({\hat K}-1/2)e_T({\hat L}-1/2)f_T({\hat T}_0-{\hat T})
g_T({\hat T}_0+{\hat T}-2) \ . 
\label{4-12b}
\end{eqnarray}
\end{subequations}
Concerning the functions (\ref{4-10a}) and (\ref{4-10b}), the inverse 
should be definable and we set up the conditions shown as follows : 
\begin{eqnarray}\label{4-13}
& &d_X(1/2)=d_X(0) \ , \quad
e_X(1/2)=e_X(0) \ , \nonumber\\
& &f_X(1)=f_X(0) \ , \qquad
g_X(1)=g_X(0) \ . \qquad\qquad
(X=S,\ T)
\end{eqnarray}
Under the above conditions, the states $\ket{c_S}$ and $\ket{c_T}$ 
can be expressed in the forms 
\begin{subequations}\label{4-14}
\begin{eqnarray}
\sqrt{\Gamma_S}\ket{c_S}&=&
\sqrt{\Gamma^0}{\hat \Omega}_S\ket{c^0}\nonumber\\
&=&\Omega_S({\hat I}{\hat J};{\hat S}{\hat S}_0)\bigl[
\ket{0}+\delta{\hat b}_-^*\ket{0}+\gamma_R\delta{\hat b}_+^*\ket{0}
+\gamma_S\delta{\hat a}_-^*\ket{0} +\gamma_R\gamma_S\delta{\hat a}_+^*\ket{0}
\nonumber\\
& &\qquad\qquad\qquad
+\gamma_T({\hat a}_+^*{\hat b}_-^*-{\hat a}_-^*{\hat b}_+^*)\ket{0}
+\cdots \bigl] \nonumber\\
&=&
\ket{0}+\delta{\hat b}_-^*\ket{0}+\gamma_R\delta{\hat b}_+^*\ket{0}
+\gamma_S\delta{\hat a}_-^*\ket{0} +\gamma_R\gamma_S\delta{\hat a}_+^*\ket{0}
\nonumber\\
& &
+\gamma_T({\hat a}_+^*{\hat b}_-^*-{\hat a}_-^*{\hat b}_+^*)\ket{0}
+\cdots \ , 
\label{4-14a}\\
\sqrt{\Gamma_T}\ket{c_T}&=&
\sqrt{\Gamma^0}{\hat \Omega}_T\ket{c^0}\nonumber\\
&=&\Omega_T({\hat K}{\hat L};{\hat T}{\hat T}_0)\bigl[
\ket{0}+\delta{\hat b}_-^*\ket{0}+\gamma_R\delta{\hat b}_+^*\ket{0}
+\gamma_S\delta{\hat a}_-^*\ket{0} +\gamma_R\gamma_S\delta{\hat a}_+^*\ket{0}
\nonumber\\
& &\qquad\qquad\qquad
+\gamma_T({\hat a}_+^*{\hat b}_-^*-{\hat a}_-^*{\hat b}_+^*)\ket{0}
+\cdots \bigl] \nonumber\\
&=&
\ket{0}+\delta{\hat b}_-^*\ket{0}+\gamma_R\delta{\hat b}_+^*\ket{0}
+\gamma_S\delta{\hat a}_-^*\ket{0} +\gamma_R\gamma_S\delta{\hat a}_+^*\ket{0}
\nonumber\\
& &
+\gamma_T({\hat a}_+^*{\hat b}_-^*-{\hat a}_-^*{\hat b}_+^*)\ket{0}
+\cdots \ .
\label{4-14b}
\end{eqnarray}
\end{subequations}
In the case where the conditions (\ref{4-13}) are not set up, 
we have the forms different from those shown in the relations 
(\ref{4-14a}) and (\ref{4-14b}). However, by scaling the parameters 
$\gamma_R$, $\gamma_S$, $\gamma_T$ and $\delta$ appropriately, 
we have the same forms as those shown in the relations (\ref{4-14a}) 
and (\ref{4-14b}) with respect to the terms expressed explicitly. 
Therefore, we treat the functions (\ref{4-10a}) and (\ref{4-10b}) 
as obeying the conditions (\ref{4-13}). In the above treatment, 
we have two types of deformations. We call the first type 
represented in $\ket{c_S}$ the $S$-type and the second 
in $\ket{c_T}$ the $T$-type deformation.

We can show that the states $\ket{c_S}$ and $\ket{c_T}$ are regarded as the 
coherent states for the operators $({\hat \gamma}_R^S, {\hat \gamma}_S^S, 
{\hat \gamma}_T^S, {\hat \delta}^S)$ and 
$({\hat \gamma}_R^T, {\hat \gamma}_S^T, {\hat \gamma}_T^T, {\hat \delta}^T)$, 
respectively : 
\begin{subequations}\label{4-15}
\begin{eqnarray}
& &{\hat \gamma}_R^S={\hat \Omega}_S{\hat \gamma}_R^0{\hat \Omega}_S^{-1}
=d_S({\hat I})e_S({\hat J}){\hat \gamma}_R^0 d_S({\hat I})^{-1}
e_S({\hat J})^{-1} \ , \nonumber\\
& &{\hat \gamma}_S^S={\hat \Omega}_S{\hat \gamma}_S^0{\hat \Omega}_S^{-1}
=f_S({\hat S}+{\hat S}_0)g_S({\hat S}-{\hat S}_0)^{-1}
{\hat \gamma}_S^0 f_S({\hat S}+{\hat S}_0)^{-1}g_S({\hat S}-{\hat S}_0) \ , 
\nonumber\\
& &{\hat \gamma}_T^S={\hat \Omega}_S{\hat \gamma}_T^0{\hat \Omega}_S^{-1}
=d_S({\hat I})e_S({\hat J}){\hat \gamma}_T^0 d_S({\hat I})^{-1}
e_S({\hat J})^{-1} \ , \nonumber\\
& &{\hat \delta}^S={\hat \Omega}_S{\hat \delta}^0{\hat \Omega}_S^{-1} \ , 
\label{4-15a}\\
& &{\hat \gamma}_R^T={\hat \Omega}_T{\hat \gamma}_R^0{\hat \Omega}_T^{-1}
=d_T({\hat K}-1/2)e_T({\hat L}-1/2){\hat \gamma}_R^0 
d_T({\hat K}-1/2)^{-1}e_T({\hat L}-1/2)^{-1} \ , \nonumber\\
& &{\hat \gamma}_S^T={\hat \Omega}_T{\hat \gamma}_S^0{\hat \Omega}_T^{-1}
=d_T({\hat K}-1/2)e_T({\hat L}-1/2){\hat \gamma}_S^0 
d_T({\hat K}-1/2)^{-1}e_T({\hat L}-1/2)^{-1} \ , \nonumber\\
& &{\hat \gamma}_T^T={\hat \Omega}_T{\hat \gamma}_T^0{\hat \Omega}_T^{-1}
=f_T({\hat T}_0-{\hat T})g_T({\hat T}_0+{\hat T}-2)
{\hat \gamma}_T^0 f_T({\hat T}_0-{\hat T})^{-1}g_T({\hat T}_0+{\hat T}-2)^{-1} 
\ , 
\nonumber\\
& &{\hat \delta}^T={\hat \Omega}_T{\hat \delta}^0{\hat \Omega}_T^{-1} \ . 
\label{4-15b}
\end{eqnarray}
\end{subequations}
Of course, we have the relations similar to those shown in the relation 
(\ref{3-12}) for $({\hat \gamma}_R^0, {\hat \gamma}_S^0, {\hat \gamma}_T^0, 
{\hat \delta}^0)$.

Finally, we should mention that the case where the functions characterizing 
the deformation have singularities can also be treated, for example, 
as was shown in Ref.\citen{9}, with the help of certain projection operators. 
In this paper, we do not contact with this problem explicitly.

\section{Construction of the $su(2)_q$- and the $su(1,1)_q$-algebras}

With the aid of the deformed boson scheme presented in the previous 
sections, we can construct the $su(2)_q$- and the $su(1,1)_q$-algebras. 
We adopt the same principle as that adopted in (III). The operators 
${\hat R}_-^0$, ${\hat S}_-^0$ and ${\hat T}_-^0$, the Hermite conjugate 
of the generators for the coherent state generating operators 
${\hat R}_+^0$, ${\hat S}_+^0$ and ${\hat T}_+^0$, are closely related 
with ${\hat \gamma}_R^0$, ${\hat \gamma}_S^0$ and ${\hat \gamma}_T^0$, 
respectively, the deformation of which is given in the relations 
(\ref{4-15a}) and (\ref{4-15b}). For ${\hat R}_-^0$, ${\hat S}_-^0$ 
and ${\hat T}_-^0$, we make the same deformation as that in the case of 
${\hat \gamma}_R^0$, ${\hat \gamma}_S^0$ and ${\hat \gamma}_T^0$ : 
\begin{subequations}\label{5-1}
\begin{eqnarray}
& &{\hat R}_+^S={\hat \Omega}_S^{-1}{\hat R}_+^0{\hat \Omega}_S \ , 
\qquad
{\hat R}_-^S={\hat \Omega}_S{\hat R}_-^0{\hat \Omega}_S^{-1} \ , \nonumber\\
& &{\hat S}_+^S={\hat \Omega}_S^{-1}{\hat S}_+^0{\hat \Omega}_S \ , 
\qquad
{\hat S}_-^S={\hat \Omega}_S{\hat S}_-^0{\hat \Omega}_S^{-1} \ , \nonumber\\
& &{\hat T}_+^S={\hat \Omega}_S^{-1}{\hat T}_+^0{\hat \Omega}_S \ , 
\qquad
{\hat T}_-^S={\hat \Omega}_S{\hat T}_-^0{\hat \Omega}_S^{-1} \ , 
\label{5-1a}\\
& &{\hat R}_+^T={\hat \Omega}_T^{-1}{\hat R}_+^0{\hat \Omega}_T \ , 
\qquad
{\hat R}_-^T={\hat \Omega}_T{\hat R}_-^0{\hat \Omega}_T^{-1} \ , \nonumber\\
& &{\hat S}_+^T={\hat \Omega}_T^{-1}{\hat S}_+^0{\hat \Omega}_T \ , 
\qquad
{\hat S}_-^T={\hat \Omega}_T{\hat S}_-^0{\hat \Omega}_T^{-1} \ , \nonumber\\
& &{\hat T}_+^T={\hat \Omega}_T^{-1}{\hat T}_+^0{\hat \Omega}_T \ , 
\qquad
{\hat T}_-^T={\hat \Omega}_T{\hat T}_-^0{\hat \Omega}_T^{-1} \ , 
\label{5-1b}
\end{eqnarray}
\end{subequations}
It should be noted that there are two types of the deformations, 
which are specified by the indices $S$ and $T$, respectively. 
Then, the deformations of $2{\hat R}_0$, $2{\hat S}_0$ and $2{\hat T}_0$ 
are given as follows : 
\begin{subequations}\label{5-2}
\begin{eqnarray}
& &[2{\hat R}_0]_S=[{\hat R}_+^S\ , \ {\hat R}_-^S]\ , \quad
[2{\hat S}_0]_S=[{\hat S}_+^S\ , \ {\hat S}_-^S]\ , \quad
[2{\hat T}_0]_S=[{\hat T}_-^S\ , \ {\hat T}_+^S]\ , 
\label{5-2a}\\
& &[2{\hat R}_0]_T=[{\hat R}_+^T\ , \ {\hat R}_-^T]\ , \quad
[2{\hat S}_0]_T=[{\hat S}_+^T\ , \ {\hat S}_-^T]\ , \quad
[2{\hat T}_0]_T=[{\hat T}_-^T\ , \ {\hat T}_+^T]\ , 
\label{5-2b}
\end{eqnarray}
\end{subequations}

In order to obtain a possible explicit expression of the forms 
(\ref{5-1}) and (\ref{5-2}), we introduce the operators ${\hat E}_X^*$ and 
${\hat E}_X\ (X=R, S, T)$ defined in the form 
\begin{eqnarray}\label{5-3}
& &{\hat E}_X^*={\hat X}_+^0\cdot \left(\sqrt{{\hat X}_-^0{\hat X}_+^0
+\epsilon}\right)^{-1}
=\left(\sqrt{{\hat X}_+^0{\hat X}_-^0+\epsilon}\right)^{-1}
\cdot {\hat X}_+^0 \ , \nonumber\\
& &{\hat E}_X={\hat X}_-^0\cdot \left(\sqrt{{\hat X}_+^0{\hat X}_-^0
+\epsilon}\right)^{-1}
=\left(\sqrt{{\hat X}_-^0{\hat X}_+^0+\epsilon}\right)^{-1}
\cdot {\hat X}_-^0 \ . \nonumber\\
& &\qquad\qquad\qquad\qquad\qquad\qquad\qquad\qquad\qquad\qquad
 (X=R, S, T)
\end{eqnarray}
Here, $\epsilon$ denotes an infinitesimal parameter. The property 
indispensable for the later discussion is given as 
\begin{equation}\label{5-4}
{\hat E}_X{\hat E}_X^*=1-\epsilon({\hat X}_-^0{\hat X}_+^0+\epsilon)^{-1} \ ,
\qquad
{\hat E}_X^*{\hat E}_X=1-\epsilon({\hat X}_+^0{\hat X}_-^0+\epsilon)^{-1} \ . 
\end{equation}
Then, for the state $\ket{\alpha}$ obeying ${\hat X}_+^0\ket{\alpha}=0$, 
we have ${\hat E}_X^*\ket{\alpha}=0$ and for $\ket{\beta}$ obeying 
${\hat X}_-^0\ket{\beta}=0$, we also have ${\hat E}_X\ket{\beta}=0$. 
Further, for the state $\ket{\gamma}$ with the conditions 
$\langle\gamma\ket{\gamma}=1$ and ${\hat X}_+^0\ket{\gamma}\neq 0$, 
the state ${\hat E}_X^*\ket{\gamma}$ is automatically normalized. 
With the use of the operators ${\hat E}_X^*$ and ${\hat E}_X$, we have 
the following formulae : 
\begin{eqnarray}\label{5-5}
{\hat X}_+^0&=&
{\hat X}_+^0\cdot\left(\sqrt{{\hat X}_-^0{\hat X}_+^0+\epsilon}\right)^{-1}
\cdot \sqrt{{\hat X}_-^0{\hat X}_+^0+\epsilon}
={\hat E}_X^*\cdot\sqrt{{\hat X}_-^0{\hat X}_+^0} 
\nonumber\\
&=&\sqrt{{\hat X}_+^0{\hat X}_-^0+\epsilon}
\cdot \left(\sqrt{{\hat X}_+^0{\hat X}_-^0+\epsilon}\right)^{-1}\cdot
{\hat X}_+^0
=\sqrt{{\hat X}_+^0{\hat X}_-^0}\cdot{\hat E}_X^* \ , \nonumber\\
{\hat X}_-^0&=&{\hat E}_X\cdot \sqrt{{\hat X}_+^0{\hat X}_-^0}
=\sqrt{{\hat X}_-^0{\hat X}_+^0}\cdot{\hat E}_X \ .
\end{eqnarray}
Further, the following relations are also useful for our aim : 
\begin{eqnarray}\label{5-6}
{\hat R}_+^0{\hat R}_-^0
&=&({\hat R}+{\hat R}_0)({\hat R}-{\hat R}_0+1)\nonumber\\
&=&({\hat S}+{\hat I}-{\hat J})({\hat S}-{\hat I}+{\hat J}+1)
=({\hat T}+{\hat K}-{\hat L})({\hat T}-{\hat K}+{\hat L}-1) \ ,
\nonumber\\
{\hat R}_-^0{\hat R}_+^0
&=&({\hat R}-{\hat R}_0)({\hat R}+{\hat R}_0+1)\nonumber\\
&=&({\hat S}-{\hat I}+{\hat J})({\hat S}+{\hat I}-{\hat J}+1)
=({\hat T}-{\hat K}+{\hat L})({\hat T}+{\hat K}-{\hat L}-1) \ ,
\nonumber\\
{\hat S}_+^0{\hat S}_-^0
&=&({\hat S}+{\hat S}_0)({\hat S}-{\hat S}_0+1)
=({\hat T}-{\hat K}-{\hat L})({\hat T}+{\hat K}+{\hat L}-1) \ ,
\nonumber\\
{\hat S}_-^0{\hat S}_+^0
&=&({\hat S}-{\hat S}_0)({\hat S}+{\hat S}_0+1)
=({\hat T}-{\hat K}-{\hat L}+1)({\hat T}+{\hat K}+{\hat L}-2) \ ,
\nonumber\\
{\hat T}_+^0{\hat T}_-^0
&=&({\hat T}_0-{\hat T})({\hat T}_0+{\hat T}-1)
=-({\hat S}-{\hat I}-{\hat J})({\hat S}+{\hat I}+{\hat J}+1) \ , 
\nonumber\\
{\hat T}_-^0{\hat T}_+^0
&=&({\hat T}_0+{\hat T})({\hat T}_0-{\hat T}+1)
=-({\hat S}-{\hat I}-{\hat J}-1)({\hat S}+{\hat I}+{\hat J}+2) \ . 
\end{eqnarray}
The use of the above relations gives us the explicit expression of the 
$S$-type deformation shown in the relations (\ref{5-1a}) and (\ref{5-2a}) : 
\begin{eqnarray}
{\hat R}_+^S&=&
{\hat E}_R^*\cdot\sqrt{[({\hat S}\!+\!1/2)+({\hat I}\!+\!1)
-({\hat J}\!+\!1/2)]_{d_Se_S}
[({\hat S}\!+\!1/2)-({\hat I}\!+\!1/2)+({\hat J})]_{d_Se_S}}\nonumber\\
&=&\sqrt{[({\hat S}\!+\!1/2)+({\hat I}\!+\!1/2)-({\hat J}\!+\!1)]_{d_Se_S}
[({\hat S}\!+\!1/2)-({\hat I})+({\hat J}\!+\!1/2)]_{d_Se_S}}
\cdot{\hat E}_R^* \ , 
\nonumber\\
& &\label{5-7a}\\
{\hat R}_-^S&=&
{\hat E}_R\cdot\sqrt{[({\hat S}\!+\!1/2)+({\hat I}\!+\!1/2)
-({\hat J}\!+\!1)]_{d_Se_S}
[({\hat S}\!+\!1/2)-({\hat I}\!+\!1/2)+({\hat J})]_{d_Se_S}}\nonumber\\
&=&\sqrt{[({\hat S}\!+\!1/2)+({\hat I}\!+\!1)-({\hat J}\!+\!1/2)]_{d_Se_S}
[({\hat S}\!+\!1/2)-({\hat I}\!+\!1/2)+({\hat J})]_{d_Se_S}}
\cdot{\hat E}_R \ , 
\nonumber\\
& &
\label{5-8a}\\
{[}2{\hat R}_0]_S&=&
[({\hat S}\!+\!1/2)+({\hat I}\!+\!1/2)-({\hat J}\!+\!1)]_{d_Se_S}
[({\hat S}\!+\!1/2)-({\hat I})+({\hat J}\!+\!1/2)]_{d_Se_S}\nonumber\\
& &-[({\hat S}\!+\!1/2)+({\hat I}\!+\!1)-({\hat J}\!+\!1/2)]_{d_Se_S}
[({\hat S}\!+\!1/2)-({\hat I}\!+\!1/2)+({\hat J})]_{d_Se_S} \ , \nonumber\\
& &
\label{5-9a}\\
{\hat S}_+^S&=&
{\hat E}_S^*\cdot\sqrt{[{\hat S}\!+\!{\hat S}_0\!+\!1]_{f_S}
[{\hat S}\!-\!{\hat S}_0]_{g_S}}
=\sqrt{[{\hat S}\!+\!{\hat S}_0]_{f_S}
[{\hat S}\!-\!{\hat S}_0\!+\!1]_{g_S}}\cdot{\hat E}_S^* \ , 
\label{5-10a}\\
{\hat S}_-^S&=&
{\hat E}_S\cdot\sqrt{[{\hat S}\!+\!{\hat S}_0]_{f_S}
[{\hat S}\!-\!{\hat S}_0\!+\!1]_{g_S}}
=\sqrt{[{\hat S}\!+\!{\hat S}_0\!+\!1]_{f_S}
[{\hat S}\!-\!{\hat S}_0]_{g_S}}\cdot{\hat E}_S \ , 
\label{5-11a}\\
{[}2{\hat S}_0]_S&=&
[{\hat S}+{\hat S}_0]_{f_S}[{\hat S}-{\hat S}_0+1]_{g_S}
-[{\hat S}+{\hat S}_0+1]_{f_S}[{\hat S}-{\hat S}_0]_{g_S} \ , 
\label{5-12a}\\
{\hat T}_+^S&=&
{\hat E}_T^*\cdot\sqrt{-[({\hat S})+({\hat I}\!+\!1)+({\hat J}\!+\!1)]_{d_Se_S}
[({\hat S})-({\hat I}\!+\!1/2)-({\hat J}\!+\!1/2)]_{d_Se_S}}\nonumber\\
&=&\sqrt{-[({\hat S})+({\hat I}\!+\!1/2)+({\hat J}\!+\!1/2)]_{d_Se_S}
[({\hat S})-({\hat I})-({\hat J})]_{d_Se_S}}\cdot{\hat E}_T^* \ , 
\label{5-13a}\\
{\hat T}_-^S&=&
{\hat E}_T\cdot\sqrt{-[({\hat S})+({\hat I}\!+\!1/2)
+({\hat J}\!+\!1/2)]_{d_Se_S}
[({\hat S})-({\hat I})-({\hat J})]_{d_Se_S}}\nonumber\\
&=&\sqrt{-[({\hat S})+({\hat I}\!+\!1)+({\hat J}\!+\!1)]_{d_Se_S}
[({\hat S})-({\hat I}\!+\!1/2)
-({\hat J}\!+\!1/2)]_{d_Se_S}}\cdot{\hat E}_T \ , \nonumber\\
& &
\label{5-14a}\\
{[}2{\hat T}_0]_S&=&
-[({\hat S})+({\hat I}\!+\!1)+({\hat J}\!+\!1)]_{d_Se_S}
[({\hat S})-({\hat I}\!+\!1/2)-({\hat J}\!+\!1/2)]_{d_Se_S}\nonumber\\
& &+[({\hat S})+({\hat I}\!+\!1/2)+({\hat J}\!+\!1/2)]_{d_Se_S}
[({\hat S})-({\hat I})-({\hat J})]_{d_Se_S} \ . 
\label{5-15a}
\end{eqnarray}
Here, $[\ \ ]_{d_Se_S}$, $[\ \ ]_{f_S}$ and $[\ \ ]_{g_S}$ are defined as 
\begin{eqnarray}
& &[(z)-p(x)-q(y)]_{d_Se_S}
=(z-px-qy)d_S(x-1/2)^{2p}e_S(y-1/2)^{2q} \ , \nonumber\\
& &
\qquad\qquad\qquad\qquad\qquad\qquad\qquad\qquad\qquad\qquad\qquad\qquad
 (p, q=\pm 1) 
\label{5-16a}\\
& &[x]_{f_S}=xf_S(x)^{-2}f_S(x-1)^2 \ , 
\label{5-17a}\\
& &[x]_{g_S}=xg_S(x)^{-2}g_S(x-1)^2 \ .
\label{5-18a}
\end{eqnarray}
For the $T$-type deformation, we have the explicit forms for the 
relations (\ref{5-1b}) and (\ref{5-2b}) : 
\begin{eqnarray}
{\hat R}_+^T&=&
{\hat E}_R^*\cdot\sqrt{[({\hat T}\!-\!1/2)+({\hat K}\!-\!1)
-({\hat L}\!-\!1/2)]_{d_Te_T}
[({\hat T}\!-\!1/2)-({\hat K}\!-\!1/2)+({\hat L})]_{d_Te_T}}\nonumber\\
&=&\sqrt{[({\hat T}\!-\!1/2)+({\hat K}\!-\!1/2)-({\hat L}\!-\!1)]_{d_Te_T}
[({\hat T}\!-\!1/2)-({\hat K})+({\hat L}\!-\!1/2)]_{d_Te_T}}
\cdot{\hat E}_R^* \ , 
\nonumber\\
& &\label{5-7b}\\
{\hat R}_-^T&=&
{\hat E}_R\cdot\sqrt{[({\hat T}\!-\!1/2)+({\hat K}\!-\!1/2)
-({\hat L}\!-\!1)]_{d_Te_T}
[({\hat T}\!-\!1/2)-({\hat K})+({\hat L}\!-\!1/2)]_{d_Te_T}}\nonumber\\
&=&\sqrt{[({\hat T}\!-\!1/2)+({\hat K}\!-\!1)-({\hat L}\!-\!1/2)]_{d_Te_T}
[({\hat T}\!-\!1/2)-({\hat K}\!-\!1/2)+({\hat L})]_{d_Te_T}}
\cdot{\hat E}_R \ , 
\nonumber\\
& &
\label{5-8b}\\
{[}2{\hat R}_0]_T&=&
[({\hat T}\!-\!1/2)+({\hat K}\!-\!1/2)-({\hat L}\!-\!1)]_{d_Te_T}
[({\hat T}\!-\!1/2)-({\hat K})+({\hat L}\!-\!1/2)]_{d_Te_T}\nonumber\\
& &-[({\hat T}\!-\!1/2)+({\hat K}\!-\!1)-({\hat L}\!-\!1/2)]_{d_Te_T}
[({\hat T}\!-\!1/2)-({\hat K}\!-\!1/2)+({\hat L})]_{d_Te_T} \ , \nonumber\\
& &
\label{5-9b}\\
{\hat S}_+^T&=&
{\hat E}_S^*\cdot\sqrt{[({\hat T})-({\hat K}\!-\!1/2)
-({\hat L}\!-\!1/2)]_{d_Te_T}
[({\hat T})+({\hat K}\!-\!1)+({\hat L}\!-\!1)]_{d_Te_T}}\nonumber\\
&=&\sqrt{[({\hat T})-({\hat K})-({\hat L})]_{d_Te_T}
[({\hat T})+({\hat K}\!-\!1/2)+({\hat L}\!-\!1/2)]_{d_Te_T}}
\cdot{\hat E}_S^* \ , 
\label{5-10b}\\
{\hat S}_-^T&=&
{\hat E}_S\cdot\sqrt{[({\hat T})-({\hat K})
-({\hat L})]_{d_Te_T}
[({\hat T})+({\hat K}\!-\!1/2)+({\hat L}\!-\!1/2)]_{d_Te_T}}\nonumber\\
&=&\sqrt{[({\hat T})-({\hat K}\!-\!1/2)-({\hat L}\!-\!1/2)]_{d_Te_T}
[({\hat T})+({\hat K}\!-\!1)
+({\hat L}\!-\!1)]_{d_Te_T}}\cdot{\hat E}_S \ , \nonumber\\
& &
\label{5-11b}\\
{[}2{\hat S}_0]_T&=&
[({\hat T})+({\hat K}\!-\!1/2)+({\hat L}\!-\!1/2)]_{d_Te_T}
[({\hat T})-({\hat K})-({\hat L})]_{d_Te_T}\nonumber\\
& &-[({\hat T})+({\hat K}\!-\!1)+({\hat L}\!-\!1)]_{d_Te_T}
[({\hat T})-({\hat K}\!-\!1/2)-({\hat L}\!-\!1/2)]_{d_Te_T} \ . 
\label{5-12b}\\
{\hat T}_+^T&=&
{\hat E}_T^*\cdot\sqrt{[{\hat T}_0\!-\!{\hat T}\!+\!1]_{f_T}
[{\hat T}_0\!+\!{\hat T}]_{g_T}}
=\sqrt{[{\hat T}_0\!-\!{\hat T}]_{f_T}
[{\hat T}_0\!+\!{\hat T}\!-\!1]_{g_T}}\cdot{\hat E}_T^* \ , 
\label{5-13b}\\
{\hat T}_-^T&=&
{\hat E}_T\cdot\sqrt{[{\hat T}_0\!-\!{\hat T}]_{f_T}
[{\hat T}_0\!+\!{\hat T}\!-\!1]_{g_T}}
=\sqrt{[{\hat T}_0\!-\!{\hat T}\!+\!1]_{f_T}
[{\hat T}_0\!+\!{\hat T}]_{g_T}}\cdot{\hat E}_T \ , 
\label{5-14b}\\
{[}2{\hat T}_0]_T&=&
[{\hat T}_0-{\hat T}+1]_{f_T}[{\hat T}_0+{\hat T}]_{g_T}
-[{\hat T}_0-{\hat T}]_{f_T}[{\hat T}_0+{\hat T}-1]_{g_T} \ , 
\label{5-15b}
\end{eqnarray}
Here, $[\ \ ]_{d_Te_T}$, $[\ \ ]_{f_T}$ and $[\ \ ]_{g_T}$ are defined as 
\begin{eqnarray}
& &[(z)-p(x)-q(y)]_{d_Te_T}
=(z-px-qy)d_T(x)^{2p}e_T(y)^{2q} \ , \nonumber\\
& &
\qquad\qquad\qquad\qquad\qquad\qquad\qquad\qquad\qquad\qquad\qquad\qquad
 (p, q=\pm 1) 
\label{5-16b}\\
& &[x]_{f_T}=xf_T(x)^{-2}f_T(x-1)^2 \ , 
\label{5-17b}\\
& &[x]_{g_T}=xg_T(x-1)^{-2}g_T(x-2)^2 \ .
\label{5-18b}
\end{eqnarray}
The above expressions are possible expressions of the $su(2)$- and 
the $su(1,1)$-algebras.

\section{An illustrative example of the application}

As a final discussion in this paper, we deal with an illustrative 
example of the applications of the method presented in this work. 
As a simple many-nucleon model, we know a shell model in which 
many-identical nucleons move in two shell-model orbits with 
the same degeneracy under the pairing interaction. 
With the use of two kinds of boson operators, this model is described 
in terms of the Holstein-Primakoff boson representation for the 
$su(2)$-algebra and we are able to observe collective motions such as 
the pairing rotation and vibration. On the other hand, the 
damped-and-amplified oscillation can be treated in the framework of the 
$su(1,1)$-algebra in the Schwinger boson representation. 
In this treatment, the extra boson operator is added to the formalism 
and it plays a role of the phase space doubling. 
The above consideration suggests us that an idea for describing 
thermal and dissipative properties of the collective motion 
observed in the above-mentioned shell model can be formulated by the 
use of four kinds of boson operators based on the framework developed 
in this paper.

First, following the idea of Ref.\citen{11,12,13}, let us derive the 
Hamiltonian in which the phase space doubling is performed for 
many-body system with two kinds of bosons $({\hat B}_+, {\hat B}_+^*)$ 
and $({\hat B}_- , {\hat B}_-^*)$. We start from the following Hamiltonian : 
\begin{subequations}\label{6-1}
\begin{eqnarray}
{\hat H}_B^0&=&
e_+{\hat B}_+^*{\hat B}_+ + e_-{\hat B}_-^*{\hat B}_-
-g({\hat B}_+^*+{\hat B}_-^*)({\hat B}_++{\hat B}_-)
\nonumber\\
& &+[g-(e_+-e_-)/2]/Z\cdot {\hat B}_+^{*2}{\hat B}_+^2
+[g+(e_+-e_-)/2]/Z\cdot {\hat B}_-^{*2}{\hat B}_-^2 
\nonumber\\
& &+(g/Z)\cdot 2{\hat B}_+^*{\hat B}_-^*{\hat B}_-{\hat B}_+ \ . 
\label{6-1a}
\end{eqnarray}
Here, $e_\pm$, $g$ and $Z$ denote real parameters characterizing the system 
under investigation and, especially, $Z$ is later 
treated as sufficiently large. Our basic idea is the same as that adopted 
in the $su(1,1)$-boson model investigated by Vitiello et. al.,\cite{11} and 
independently by the present authors.\cite{12} 
For the present aim, we introduce the boson operators 
$({\hat A}_+ , {\hat A}_+^*)$ and $({\hat A}_- , {\hat A}_-^*)$ for 
the phase space doubling. The Hamiltonian for this boson system is 
obtained by replacing $({\hat B}_\pm , {\hat B}_\pm^*)$ with 
$({\hat A}_\pm , {\hat A}_\pm^*)$ for the Hamiltonian (\ref{6-1a}) : 
\begin{eqnarray}
{\hat H}_A^0&=&
e_+{\hat A}_+^*{\hat A}_+ + e_-{\hat A}_-^*{\hat A}_-
-g({\hat A}_+^*+{\hat A}_-^*)({\hat A}_++{\hat A}_-)
\nonumber\\
& &+[g-(e_+-e_-)/2]/Z\cdot {\hat A}_+^{*2}{\hat A}_+^2
+[g+(e_+-e_-)/2]/Z\cdot {\hat A}_-^{*2}{\hat A}_-^2 
\nonumber\\
& &+(g/Z)\cdot 2{\hat A}_+^*{\hat A}_-^*{\hat A}_-{\hat A}_+ \ . 
\label{6-1b}
\end{eqnarray}
Further, ${\hat H}_I^0$ is introduced for the interaction between both 
kinds of bosons : 
\begin{eqnarray}
{\hat H}_I^0&=&
i\chi[({\hat A}_+^*{\hat B}_+^*+{\hat A}_-^*{\hat B}_-^*)-({\hat B}_+
{\hat A}_++{\hat B}_-{\hat A}_-)]\nonumber\\
& &+(e_+-e_-)/Z\cdot ({\hat A}_-^*{\hat B}_+^*{\hat B}_+{\hat A}_-
-{\hat A}_+^*{\hat B}_-^*{\hat B}_-{\hat A}_+) \ .
\label{6-1c}
\end{eqnarray}
\end{subequations}
Then, the Hamiltonian in the phase space doubling, ${\hat H}^0$, is 
expressed in the form 
\begin{equation}\label{6-2}
{\hat H}^0={\hat H}_B^0-{\hat H}_A^0+{\hat H}_I^0 \ .
\end{equation}
The Hamiltonian (\ref{6-2}) can be rewritten as 
\begin{eqnarray}\label{6-3}
{\hat H}^0&=&
-(e_++e_-)/2\cdot 2{\hat S}_0+(e_+-e_-)/2\cdot 2{\hat R}_0
[1+(2{\hat T}_0-3)/Z]\nonumber\\
& &+g\cdot 2{\hat S}_0[1-(2{\hat T}_0-3)/Z]-g({\hat R}_+^0+{\hat R}_-^0)
+i\chi({\hat T}_+^0-{\hat T}_-^0) \ .
\end{eqnarray}
The above expression is obtained by reading $({\hat B}_\pm , {\hat A}_\pm)$ 
as 
\begin{equation}\label{6-4}
{\hat B}_+ \rightarrow {\hat b}_+ \ , \quad
{\hat B}_- \rightarrow {\hat b}_- \ , \quad
{\hat A}_+ \rightarrow -{\hat a}_- \ , \quad
{\hat A}_- \rightarrow {\hat a}_+ \ .
\end{equation}
Then, the deformed Hamiltonian ${\hat H}^X\ (X=S, T)$ is written down 
in the form 
\begin{eqnarray}\label{6-5}
{\hat H}^X&=&
-(e_++e_-)/2\cdot [2{\hat S}_0]_X+(e_+-e_-)/2\cdot [2{\hat R}_0]_X
[1+([2{\hat T}_0]_X-3)/Z]\nonumber\\
& &+g\cdot [2{\hat S}_0]_X[1-([2{\hat T}_0]_X-3)/Z]
-g\cdot({\hat R}_+^X+{\hat R}_-^X)
+i\chi\cdot({\hat T}_+^X-{\hat T}_-^X) \ . \nonumber\\
& &\qquad\qquad\qquad\qquad\qquad\qquad\qquad\qquad\qquad\qquad 
(X=S, T)
\end{eqnarray}

In this paper, we investigate the $S$-type deformation concretely 
under the following condition :
\begin{eqnarray}\label{6-6}
& &f_S({\hat S}+{\hat S}_0)=g_S({\hat S}-{\hat S}_0)=1 \ , \nonumber\\
& &d_S({\hat I}+1/2)^{-1}d_S({\hat I})=\sqrt{1-2{\hat I}/Z} \ , 
\quad
e_S({\hat J})^{-1}e_S({\hat J}+1/2)=\sqrt{1-2{\hat J}/Z} \ . \qquad
\end{eqnarray}
We have already investigated the effects of $f_S$, $g_S$, $f_T$ and $g_T$ 
in (II). In the present paper, the functions $d_S$, $e_S$, $d_T$ and $e_T$ 
are newly added. Therefore, in order to investigate the effects 
coming from $d_S$, $e_S$, $d_T$ and $e_T$, we adopt the condition (\ref{6-6}). 
Then, ${\hat H}^X$ for $X=S$ given in the relation (\ref{6-5}) can be 
expressed as 
\begin{equation}\label{6-7}
{\hat H}^S={\hat H}_B^S-{\hat H}_A^S+{\hat H}_I^S+{\hat h}_B^S-{\hat h}_A^S \ .
\end{equation}
Here, each term is given as 
\begin{subequations}\label{6-8}
\begin{eqnarray}
{\hat H}_B^S&=&
e_+{\hat B}_+^*{\hat B}_+ + e_-{\hat B}_-^*{\hat B}_- \nonumber\\
& &-(g/Z)\cdot\left({\hat B}_+^*\sqrt{Z-{\hat A}_-^*{\hat A}_-
-{\hat B}_+^*{\hat B}_+} + {\hat B}_-^*\sqrt{Z-{\hat A}_+^*{\hat A}_+
-{\hat B}_-^*{\hat B}_-}\right) \nonumber\\
& &\quad\quad \times
\left(\sqrt{Z-{\hat A}_-^*{\hat A}_-
-{\hat B}_+^*{\hat B}_+}\ {\hat B}_+ 
+ \sqrt{Z-{\hat A}_+^*{\hat A}_+ -{\hat B}_-^*{\hat B}_-}\ {\hat B}_-
\right) \ , 
\qquad\quad
\label{6-8a}\\
{\hat H}_A^S&=&
e_+{\hat A}_+^*{\hat A}_+ + e_-{\hat A}_-^*{\hat A}_- \nonumber\\
& &-(g/Z)\cdot\left({\hat A}_+^*\sqrt{Z-{\hat B}_-^*{\hat B}_-
-{\hat A}_+^*{\hat A}_+} + {\hat A}_-^*\sqrt{Z-{\hat B}_+^*{\hat B}_+
-{\hat A}_-^*{\hat A}_-}\right) \nonumber\\
& &\quad\quad \times
\left(\sqrt{Z-{\hat B}_-^*{\hat B}_-
-{\hat A}_+^*{\hat A}_+}\ {\hat A}_+ 
+ \sqrt{Z-{\hat B}_+^*{\hat B}_+ -{\hat A}_-^*{\hat A}_-}\ {\hat A}_-
\right) \ , 
\label{6-8b}\\
{\hat H}_I^S&=&
(i\chi/Z)\cdot ({\hat A}_+^*\sqrt{Z-{\hat B}_-^*{\hat B}_-
-{\hat A}_+^*{\hat A}_+}\cdot{\hat B}_+^*\sqrt{Z-{\hat A}_-^*{\hat A}_-
-{\hat B}_+^*{\hat B}_+} \nonumber\\
& &\qquad\qquad +
{\hat A}_-^*\sqrt{Z-{\hat B}_+^*{\hat B}_+
-{\hat A}_-^*{\hat A}_-}\cdot{\hat B}_-^*\sqrt{Z-{\hat A}_+^*{\hat A}_+
-{\hat B}_-^*{\hat B}_-}\nonumber\\
& &\qquad\qquad -
\sqrt{Z-{\hat B}_-^*{\hat B}_- -{\hat A}_+^*{\hat A}_+}\ {\hat A}_+ \cdot
\sqrt{Z-{\hat A}_-^*{\hat A}_- -{\hat B}_+^*{\hat B}_+}\ {\hat B}_+ 
\nonumber\\
& &\qquad\qquad -
\sqrt{Z-{\hat B}_+^*{\hat B}_+ -{\hat A}_-^*{\hat A}_-}\ {\hat A}_- \cdot
\sqrt{Z-{\hat A}_+^*{\hat A}_+ -{\hat B}_-^*{\hat B}_-}\ {\hat B}_-), \ \qquad
\label{6-8c}
\end{eqnarray}
\end{subequations}
\vspace{-0.8cm}
\begin{subequations}\label{6-9}
\begin{eqnarray}
& &{\hat h}_B^S=(g/Z)\cdot 2{\hat B}_+^*{\hat B}_-^*{\hat B}_-{\hat B}_+ \ , 
\qquad\qquad\qquad\qquad\qquad\qquad\qquad\qquad\qquad
\label{6-9a}\\
& &{\hat h}_A^S=(g/Z)\cdot 2{\hat A}_+^*{\hat A}_-^*{\hat A}_-{\hat A}_+ \ . 
\label{6-9b}
\end{eqnarray}
\end{subequations}
The above is the Hamiltonian deformed from the Hamiltonian (\ref{6-2}) for 
the case (\ref{6-6}). It should be noted that the Hamiltonian (\ref{6-7}) 
is derived approximately under the condition that $Z$ is sufficiently large. 
Later, the meaning of this approximation is mentioned.

It is now possible to investigate the relation between the above-derived 
Hamiltonian and that for the conventional pairing correlation 
in two-level shell model. For this nucleon system, we are able to write 
down the Hamiltonian in the form 
\begin{eqnarray}\label{6-10}
{\maru H}&=&(\epsilon_+-\lambda)\Omega+(\epsilon_--\lambda)\Omega 
+2(\epsilon_+-\lambda){\maru S}_0(+)+2(\epsilon_--\lambda){\maru S}_0(-) 
\nonumber\\
& &\qquad -G({\maru S}{}_+^0(+)+{\maru S}{}_+^0(-))({\maru S}{}_-^0(+)
+{\maru S}{}_-^0(-)) \ .
\end{eqnarray}
Here, the indices $+$ and $-$ denote the upper and the lower level, 
the degeneracies of which are the same as each other ($\Omega=(2j+1)/2\ : \ 
j=$the spin of the $j$-$j$ coupling shell model state). 
The parameters $\epsilon_\pm$, $\lambda$ and $G$ mean the 
single-particle energies, the chemical potential and the strength of 
the pairing interaction. The sets 
$({\maru S}{}_\pm^0(+) , {\maru S}_0(+))$ and 
$({\maru S}{}_\pm^0(-) , {\maru S}_0(-))$ form the $su(2)$-algebras 
for the levels $\pm$, respectively. 
The Holstein-Primakoff boson representation for ${\maru S}{}_+^0(\pm)$, 
${\maru S}{}_-^0(\pm)$ and ${\maru S}_0(\pm)$ are given by 
\begin{eqnarray}\label{6-11a}
& &{\maru S}{}_+^0(B_\pm)={\maru B}{}_\pm^*\sqrt{\Omega-{\maru \nu}(B_\pm)
-{\maru B}{}_\pm^*{\maru B}_\pm} \ , \nonumber\\
& &{\maru S}{}_-^0(B_\pm)=\sqrt{\Omega-{\maru \nu}(B_\pm)
-{\maru B}{}_\pm^*{\maru B}_\pm}\ {\maru B}_\pm \ , \nonumber\\
& &{\maru S}_0(B_\pm)=-(\Omega-{\maru \nu}(B_\pm))/2+
{\maru B}{}_\pm^*{\maru B}_\pm \ . 
\end{eqnarray}
Here, $({\maru B}_\pm , {\maru B}{}_\pm^*)$ denote boson operators. The 
quantities ${\maru \nu}(B_\pm)$ play a role of the seniority numbers 
of the levels $\pm$ and they are independent of the bosons 
$({\maru B}_\pm , {\maru B}{}_\pm^*)$. 
Then, the Hamiltonian (\ref{6-10}) can be rewritten as 
\begin{eqnarray}\label{6-12a}
{\maru H}_B&=&(\epsilon_+-\lambda)({\maru \nu}(B_+)+2{\maru B}{}_+^*
{\maru B}_+)+(\epsilon_--\lambda)({\maru \nu}(B_-)+2{\maru B}{}_-^*
{\maru B}_-)\nonumber\\
& &
-G\left({\maru B}{}_+^*\sqrt{\Omega-{\maru \nu}(B_+)-{\maru B}{}_+^*
{\maru B}_+}+{\maru B}{}_-^*\sqrt{\Omega-{\maru \nu}(B_-)
-{\maru B}{}_-^*{\maru B}_-}\right)\nonumber\\
& &\quad\times 
\left(\sqrt{\Omega-{\maru \nu}(B_+)-{\maru B}{}_+^*{\maru B}_+}\ 
{\maru B}_+ +\sqrt{\Omega-{\maru \nu}(B_-)
-{\maru B}{}_-^*{\maru B}_-}\ {\maru B}_- \right) .\qquad
\end{eqnarray}
With the aim of the phase space doubling, we, further, introduce 
the operators ${\maru S}{}_+^0(A_\pm)$, ${\maru S}{}_-^0(A_\pm)$ and 
${\maru S}_0(A_\pm)$ in the form 
\begin{eqnarray}\label{6-11b}
& &{\maru S}{}_+^0(A_\pm)={\maru A}{}_\pm^*\sqrt{\Omega-{\maru \nu}(A_\pm)
-{\maru A}{}_\pm^*{\maru A}_\pm} \ , \nonumber\\
& &{\maru S}{}_-^0(A_\pm)=\sqrt{\Omega-{\maru \nu}(A_\pm)
-{\maru A}{}_\pm^*{\maru A}_\pm}\ {\maru A}_\pm \ , \nonumber\\
& &{\maru S}_0(A_\pm)=-(\Omega-{\maru \nu}(A_\pm))/2+
{\maru A}{}_\pm^*{\maru A}_\pm \ . 
\end{eqnarray}
The Hamiltonian ${\maru H}_A$ which is necessary for the phase space doubling 
is given in the form 
\begin{eqnarray}\label{6-12b}
{\maru H}_A&=&(\epsilon_+-\lambda)({\maru \nu}(A_+)+2{\maru A}{}_+^*
{\maru A}_+)+(\epsilon_--\lambda)({\maru \nu}(A_-)+2{\maru A}{}_-^*
{\maru A}_-)\nonumber\\
& &
-G\left({\maru A}{}_+^*\sqrt{\Omega-{\maru \nu}(A_+)-{\maru A}{}_+^*
{\maru A}_+}+{\maru A}{}_-^*\sqrt{\Omega-{\maru \nu}(A_-)
-{\maru A}{}_-^*{\maru A}_-}\right)\nonumber\\
& &\quad\times 
\left(\sqrt{\Omega-{\maru \nu}(A_+)-{\maru A}{}_+^*{\maru A}_+}\ 
{\maru A}_+ +\sqrt{\Omega-{\maru \nu}(A_-)
-{\maru A}{}_-^*{\maru A}_-}\ {\maru A}_- \right) .\qquad
\end{eqnarray}
The meaning of the notations is self-evident. Under the correspondence 
$({\maru B}_\pm \sim {\hat B}_\pm \ , \ {\maru A}_\pm \sim {\hat A}_\pm)$, 
we can see that $({\maru H}_B-{\maru H}_A)$ is reduced to 
$({\hat H}_B-{\hat H}_A)$ shown in the relations (\ref{6-8a}) 
and (\ref{6-8b}), if the following relations hold : 
\begin{eqnarray}\label{6-13}
& &e_+=2(\epsilon_+ - \lambda)-(\epsilon_- - \lambda) \ , \qquad
e_-=2(\epsilon_- - \lambda)-(\epsilon_+ - \lambda) \ , \nonumber\\
& &g=G\Omega \ , \qquad\qquad Z=\Omega \ .
\end{eqnarray}
In this case, the seniority numbers ${\maru \nu}(B_\pm)$ and 
${\maru \nu}(A_\pm)$ are obtained in the form 
\begin{equation}\label{6-14}
{\maru \nu}(B_+)={\maru A}{}_-^*{\maru A}_- \ , \quad
{\maru \nu}(B_-)={\maru A}{}_+^*{\maru A}_+ \ , \quad
{\maru \nu}(A_+)={\maru B}{}_-^*{\maru B}_- \ , \quad
{\maru \nu}(A_-)={\maru B}{}_+^*{\maru B}_+ \ .
\end{equation}
Then, it may be permitted to adopt the following form as the 
interaction part : 
\begin{eqnarray}\label{6-15}
{\maru H}_I&=&(i\chi/\Omega)\cdot
\bigl({\maru S}{}_+^0(A_+){\maru S}{}_+^0(B_+)
+{\maru S}{}_+^0(A_-){\maru S}{}_+^0(B_-)\nonumber\\
& &\qquad\qquad 
-{\maru S}{}_-^0(B_+){\maru S}{}_-^0(A_+)
-{\maru S}{}_-^0(B_-){\maru S}{}_-^0(A_-)\bigl) \ .
\end{eqnarray}
However, it should be noted that in the two-level model, 
there does not exist the term such as $({\hat h}_B^S-{\hat h}_A^S)$ 
shown in the relation (\ref{6-9}).

The above is an illustrative example. For the above treatment, we 
give two comments. First is related to a possible introduction of the 
seniority numbers. 
For example, the forms (\ref{6-11a}), (\ref{6-11b}) 
and (\ref{6-14}) tell us that 
${\maru A}{}_-^*{\maru A}_-$ plays a role of the seniority numbers 
${\maru \nu}(B_+)$ for the pairing correlation described by the boson 
$({\maru B}_+ , {\maru B}{}_+^*)$. We know that if ${\maru \nu}(B_+)$ 
becomes larger, the pairing correlation becomes weaker and, if 
${\maru A}{}_-^*{\maru A}_-$ becomes larger, the statistical mixing 
becomes larger. The above suggests us that if the statistical mixing 
becomes stronger, the pairing correlation becomes weaker. This may be 
acceptable as a natural feature. This is the first comment. 
Second is related with ${\maru H}_I$ shown in the relation (\ref{6-15}). 
In the case of an example of the deformation for two kinds of bosons, 
we got such a term $({\maru S}{}_+^0(a){\maru S}{}_+^0(b)
-{\maru S}{}_-^0(b){\maru S}{}_-^0(a))$. 
The form (\ref{6-15}) is its natural generalization. Therefore, 
it may be interesting to investigate these points concretely, 
together with the term $({\hat h}_B^S-{\hat h}_A^S)$.

\section{Concluding remarks}

In this paper, we proposed a possible deformed boson scheme for 
many-body systems consisting of four kinds of boson operators. 
The basic viewpoint was to focus the interest on the 
$su(2)$- and the $su(1,1)$-algebras. As an example of the application, 
we investigated two-level shell model under the pairing correlation 
in the framework of the deformed boson scheme under a special condition. 
In the present section, as the concluding remarks, we discuss the scheme 
given in \S 6 in the language of the $su(3)\times su(3)$-algebra 
in a symmetric representation presented in Ref.\citen{16}. 
For this aim, we prepare three kinds of boson operators 
$({\hat \xi} , {\hat \xi}^*)$, $({\hat \eta} , {\hat \eta}^*)$ and 
$({\hat \zeta} , {\hat \zeta}^*)$ and define the following 
bilinear form : 
\begin{subequations}\label{7-1}
\begin{eqnarray}
& &{\hat S}_+^0={\hat \xi}^*{\hat \zeta} \ ,\quad
{\hat S}_-^0={\hat \zeta}^*{\hat \xi} \ , \quad
{\hat S}_0=({\hat \xi}^*{\hat \xi}-{\hat \zeta}^*{\hat \zeta})/2 \ , 
\label{7-1a}\\
& &{\hat \Sigma}_+^0={\hat \eta}^*{\hat \zeta} \ ,\quad
{\hat \Sigma}_-^0={\hat \zeta}^*{\hat \eta} \ , \quad
{\hat \Sigma}_0=({\hat \eta}^*{\hat \eta}-{\hat \zeta}^*{\hat \zeta})/2 \ , 
\label{7-1b}\\
& &{\hat \sigma}_+^0={\hat \xi}^*{\hat \eta} \ ,\quad
{\hat \sigma}_-^0={\hat \eta}^*{\hat \xi} \ , \quad
{\hat \sigma}_0=({\hat \xi}^*{\hat \xi}-{\hat \eta}^*{\hat \eta})/2 \ . 
\label{7-1c}
\end{eqnarray}
\end{subequations}
The sets $({\hat S}_\pm^0 , {\hat S}_0)$, 
$({\hat \Sigma}_\pm^0 , {\hat \Sigma}_0)$ and 
$({\hat \sigma}_\pm^0 , {\hat \sigma}_0)$ form the $su(2)$-algebras. 
However, since ${\hat \sigma}_0={\hat S}_0-{\hat \Sigma}_0$, 
$({\hat S}_\pm^0 , {\hat S}_0)$, 
$({\hat \Sigma}_\pm^0 , {\hat \Sigma}_0)$ and 
$({\hat \sigma}_\pm^0)$ are independent of each other. 
These operators form the $su(3)$-algebra in the symmetric representation 
and we can call the set the Schwinger boson representation.

We can prove that all the generators commute with the following operator : 
\begin{equation}\label{7-2}
{\hat K}=({\hat \xi}^*{\hat \xi}+{\hat \eta}^*{\hat \eta}
+{\hat \zeta}^*{\hat \zeta})/2 \ .
\end{equation}
Then, the Holstein-Primakoff boson representation for a fixed total boson 
number can be constructed and the result is as follows : 
\begin{subequations}\label{7-3}
\begin{eqnarray}
& &{\maru S}{}_+^0 = {\maru \xi}{}^* 
\sqrt{2K-{\maru \eta}{}^*{\maru \eta}-{\maru \xi}{}^*{\maru \xi}} \ , 
\qquad
{\maru S}{}_-^0 = 
\sqrt{2K-{\maru \eta}{}^*{\maru \eta}-{\maru \xi}{}^*{\maru \xi}}\ 
{\maru \xi} \ , \nonumber\\
& &{\maru S}_0=-(K-{\maru \eta}{}^*{\maru \eta}/2)
+{\maru \xi}{}^*{\maru \xi} \ , 
\label{7-3a}\\
& &{\maru \Sigma}{}_+^0 = {\maru \eta}{}^* 
\sqrt{2K-{\maru \xi}{}^*{\maru \xi}-{\maru \eta}{}^*{\maru \eta}} \ , 
\qquad
{\maru \Sigma}{}_-^0 = 
\sqrt{2K-{\maru \xi}{}^*{\maru \xi}-{\maru \eta}{}^*{\maru \eta}}\ 
{\maru \eta} \ , \nonumber\\
& &{\maru \Sigma}_0=-(K-{\maru \xi}{}^*{\maru \xi}/2)
+{\maru \eta}{}^*{\maru \eta} \ , 
\label{7-3b}\\
& &{\maru \sigma}{}_+^0 = {\maru \xi}{}^* {\maru \eta} \ , 
\qquad
{\maru \sigma}{}_-^0 = {\maru \eta}{}^*{\maru \xi} \ , \qquad
{\maru \sigma}_0=({\maru \xi}{}^*{\maru \xi}-
{\maru \eta}{}^*{\maru \eta})/2 \ . 
\label{7-3c}
\end{eqnarray}
\end{subequations}
Here, $({\maru \xi} , {\maru \xi}{}^*)$ and 
$({\maru \eta} , {\maru \eta}{}^*)$ denote the boson operators. 
The quantity $K$ is the eigenvalue of ${\hat K}$ defined in the relation 
(\ref{7-2}) and $K=0, 1/2, 1, 3/2, \cdots$. 
Judging from the form of the relation (\ref{7-3}), we can call the set 
(\ref{7-3}) as the Holstein-Primakoff boson representation for 
the $su(3)$-algebra in the symmetric representation.

On the basis of the relations (\ref{7-3a}) and (\ref{7-3b}), 
let us reinvestigate the forms (\ref{6-11a}) and (\ref{6-11b}). 
In this case, the relation (\ref{6-14}) is also noted. 
In associating with the forms (\ref{6-11a}) and (\ref{6-11b}), we define 
the operator 
\begin{eqnarray}\label{7-4}
& &{\maru \sigma}{}_+^0(+)={\maru B}{}_+^*{\maru A}_- \ , \qquad
{\maru \sigma}{}_-^0(+)={\maru A}{}_-^*{\maru B}_+ \ , \nonumber\\
& &{\maru \sigma}{}_+^0(-)={\maru B}{}_-^*{\maru A}_+ \ , \qquad
{\maru \sigma}{}_-^0(-)={\maru A}{}_+^*{\maru B}_- \ . 
\end{eqnarray}
Together with the form (\ref{7-4}), we can see that the sets 
$({\maru S}{}_\pm^0(B_+) , {\maru S}_0(B_+) , 
{\maru S}{}_\pm^0(A_-) ,$\break
${\maru S}_0(A_-) , {\maru \sigma}{}_\pm^0(+))$ 
and $({\maru S}{}_\pm^0(B_-) , {\maru S}_0(B_-) , 
{\maru S}{}_\pm^0(A_+) , {\maru S}_0(A_+) , {\maru \sigma}{}_\pm^0(-))$ 
form two independent Holstein-Primakoff boson representations of the 
$su(3)$-algebras in the symmetric representation. 
The meaning of $\Omega$ may be self-evident and in the shell model 
adopted in this paper, usually, $\Omega$ is regarded as sufficiently large. 
Therefore, from the relation (\ref{6-13}), we can regard $Z$ as 
sufficiently large. From the above argument, we are able to obtain 
the Holstein-Primakoff boson representation of the 
$su(3)\times su(3)$-algebra in the symmetric representation 
as a result of a possible deformed boson scheme for four kinds of 
boson operators.


\appendix
\section{The proof of the explicit expression of the operator ${\hat T}$}

First, let us sketch the eigenvalue problem for the $su(1,1)$-algebra 
in the systems described in terms of two and four kinds of boson operators. 
We presuppose the existence of a state $\ket{t}$, which obeys 
\begin{equation}\label{A-1}
{\hat T}_-^0\ket{t}=0 \ ,\qquad
{\hat {\mib T}}^2\ket{t}=t(t-1)\ket{t} \ , \qquad
{\hat T}_0\ket{t}=\tau\ket{t} \ .
\end{equation}
Here, $\tau$ is a function of $t$, the explicit form of which should be 
determined in the framework of the treatment. In this Appendix, there 
appear various states in which we omit the normalization constants. 
Further, for a moment, we omit the extra-quantum numbers additional 
to $t$. Since the Casimir operator can be expressed in the form 
${\hat {\mib T}}^2={\hat T}_0({\hat T}_0-1)-{\hat T}_+^0{\hat T}_-^0$, 
the relation (\ref{A-1}) gives us 
\begin{equation}\label{A-2}
\tau(\tau-1)=t(t-1) \ .
\end{equation}
As a formal solution of Eq.(\ref{A-2}), we have 
\begin{eqnarray}
& &t=\tau \ , 
\label{A-3}\\
& &t=1-\tau \ . 
\label{A-4}
\end{eqnarray}
The above means that $\tau$ can be expressed in terms of $t$.

For the case of two kinds of boson operators, the relation (\ref{2-1b}) 
gives us 
\begin{equation}\label{A-5}
\tau=1/2+n/2 \ . \qquad (n=0, 1, 2, \cdots)
\end{equation}
From the relations (\ref{A-3})$\sim$(\ref{A-5}), the following forms 
are derived : 
\begin{eqnarray}
& &t=1/2+n/2 \ . \qquad (t=1/2,\ 1,\ 3/2, \cdots) 
\label{A-6}\\
& &t=1/2-n/2 \ . \qquad (t=1/2,\ 0,\ -1/2 , \cdots)
\label{A-7}
\end{eqnarray}
The case $n=0$ corresponds to the boson vacuum $\ket{0}$ and two 
expressions (\ref{A-6}) and (\ref{A-7}) coincide with each other $(t=1/2)$. 
Then, $\ket{t=1/2}$ can be expressed as 
\begin{equation}\label{A-8}
\ket{t=1/2}=\ket{0} \ .
\end{equation}
In the case $n=1, 2, 3, \cdots$, we make the relations (\ref{A-6}) 
and (\ref{A-7}) correspond to the following states, respectively : 
\begin{eqnarray}
& &\ket{t=1/2+n/2}=(\sqrt{n!})^{-1}({\hat b}^*)^n \ket{0} \ , 
\label{A-9}\\
& &\ket{t=1/2-n/2}=(\sqrt{n!})^{-1}({\hat a}^*)^n \ket{0} \ . 
\label{A-10}
\end{eqnarray}
The relation (\ref{A-6}) and (\ref{A-7}) can be unified into 
the form $n/2=|t-1/2|$ and, with the aid of the relation (\ref{A-5}), 
we have 
\begin{equation}\label{A-11}
\tau=1/2+|t-1/2| \ . \qquad (t=1/2,\ 1,\ 3/2,\ \cdots )
\end{equation}
Successive operations of ${\hat T}_+^0$ on the state $\ket{t}$ 
gives us the eigenstates of ${\hat {\mib T}}^2$ and ${\hat T}_0$ 
in the form 
\begin{eqnarray}
& &\ket{t,t_0}=({\hat T}_+^0)^{t_0-(1/2+|t-1/2|)} \ket{t} \ , 
\label{A-12}\\
& & \qquad t_0=1/2+|t-1/2|, \ 3/2+|t-1/2| , \ 5/2+|t-1/2| \ ,\cdots \ .
\label{A-13}
\end{eqnarray}
Of course, $t_0$ denotes the eigenvalue of ${\hat T}_0$. 
The relations (\ref{A-6}), (\ref{A-7}), (\ref{A-9}) and (\ref{A-10}) 
suggest us 
\begin{equation}\label{A-14}
{\hat {\mib T}}^2={\hat T}({\hat T}-1) \ , \qquad
{\hat T}=1/2+({\hat N}_b-{\hat N}_a)/2 \ . 
\end{equation}
The expression (\ref{A-14}) is nothing but the form introduced in \S 2.

For the case of four kinds of boson operators, we have 
\begin{equation}\label{A-15}
\tau=1+n/2 \ . \qquad (n=0, 1, 2, \cdots )
\end{equation}
Judging from the expression (\ref{3-1b}), the above form may be 
accepted as self-evident. From the relations (\ref{A-3}), (\ref{A-4}) 
and (\ref{A-15}), the following forms are derived : 
\begin{eqnarray}
& &t=1+n/2 \ , \qquad (t=1,\ 3/2,\ 2, \cdots) 
\label{A-16}\\
& &t=-n/2 \ , \quad\qquad (t=0,\ -1/2,\ -1, \cdots) 
\label{A-17}
\end{eqnarray}
The case $n=0$ corresponds to the boson vacuum $\ket{0}$ and two expressions 
(\ref{A-16}) and (\ref{A-17}) do not coincide with each other : 
\begin{eqnarray}
& &\ket{t=1}=\ket{0} \ , 
\label{A-18}\\
& &\ket{t=0}=\ket{0} \ . 
\label{A-19}
\end{eqnarray}
This situation is quite different from the case (\ref{A-8}). 
This case is attributed to the manner how to specify the vacuum $\ket{0}$ 
in the present quantum number. There exist two manners : 
the form (\ref{A-18}) or (\ref{A-19}). In the present, we adopt the form 
(\ref{A-18}). Therefore, as the continuation from the form (\ref{A-18}), 
it may be natural to adopt the form (\ref{A-16}) for $n=1,2,3,\cdots$ : 
\begin{equation}\label{A-20}
\ket{t=1+n/2}=({\hat b}_-^*)^n \ket{0} \ (=\ket{n}) \ .
\end{equation}
However, in the present case, it is noted that the state which satisfies 
the relation (\ref{A-1}) is not restricted to the form (\ref{A-20}). 
By successive operations of ${\hat S}_+^0$ and ${\hat R}_+^0$ on the 
state $\ket{t=1+n/2}$, we have 
\begin{equation}\label{A-21}
\ket{m_S\ m_R : t=1+n/2}=({\hat S}_+^0)^{m_S}({\hat R}_+^0)^{m_R}
\ket{n}\ (=\ket{m_S\ m_R :t})\ .
\end{equation}
In the above, $m_S$ and $m_R$ denote the extra-quantum numbers additional 
to $t$. The state $\ket{m_S\ m_R : t}$ is the eigenstate of 
${\hat S}_0$ and ${\hat R}_0$ with the eigenvalues $m_S+n/2$ and 
$m_R+n/2$, respectively. 
By successive operation of the operator ${\hat T}_+^0$ on the state 
(\ref{A-21}), we have the eigenstate of ${\hat {\mib T}}^2$ and 
${\hat T}_0$ : 
\begin{eqnarray}
& &\ket{m_S\ m_R :t,t_0}=({\hat T}_+^0)^{t_0-t} \ket{m_S\ m_R :t} \ , 
\label{A-22}\\
& &\qquad t_0=t,\ t+1,\ t+2,\cdots \ .
\label{A-23}
\end{eqnarray}
Hereafter, concerning the quantum numbers, we discuss only $t$ and $t_0$ 
and, then, we omit the quantum numbers $m_S$ and $m_R$. The state 
$\ket{t,t_0}$ satisfies the relations 
\begin{eqnarray}
& &({\hat T}_-^0)^r\ket{t,t_0}
=\sqrt{(t_0\!-\!t)!(t_0\!+\!t\!-\!1)!}
\left(\sqrt{(t_0\!-\!t\!-\!r)!(t_0\!+\!t\!-\!r\!-\!1)!}\right)^{-1}
\ket{t, t_0-r} \ , \nonumber\\
& &\label{A-24}\\
& &({\hat T}_+^0)^r\ket{t,t_0}
=\left(\sqrt{(t_0\!-\!t)!(t_0\!+\!t\!-\!1)!}\right)^{-1}
\sqrt{(t_0\!-\!t\!+\!r)!(t_0\!+\!t\!+\!r\!-\!1)!}
\ket{t, t_0+r} \ . \nonumber\\
& &\label{A-25}
\end{eqnarray}
In the case of two kinds of bosons, we have the operator ${\hat T}$ 
shown in Eq.(\ref{A-14}). The aim of this Appendix is to give the explicit 
form of ${\hat T}$ for the case of four kinds of boson operators.

If the Casimir operator of the $su(1,1)$-algebra, ${\hat {\mib T}}^2$, 
can be expressed in the form ${\hat {\mib T}}^2={\hat T}({\hat T}-1)$, 
${\hat T}$ satisfies 
\begin{equation}\label{A-26}
{\hat T}\ket{t,t_0}=t\ket{t,t_0} \ .
\end{equation}
In order to search the explicit form of ${\hat T}$ itself, we introduce 
the following operator ; 
\begin{eqnarray}
& &{\hat \tau}_0=\sum_{r_0=1}^{\infty}
g_{r_0}({\hat T}_+^0)^{r_0}({\hat Z}_{r_0})^{-1}({\hat T}_-^0)^{r_0} \ ,
\label{A-27}\\
& &{\hat Z}_{r_0}=(2{\hat T}_0)(2{\hat T}_0+1)\cdots(2{\hat T}_0+2r_0-2) \ .
\label{A-28}
\end{eqnarray}
Since ${\hat T}_0$ is positive-definite and $({\hat Z}_{r_0})^{-1}$ 
is sandwiched between $({\hat T}_+^0)^{r_0}$ and 
$({\hat T}_-^0)^{r_0}$, the operator ${\hat \tau}_0$ can be defined. 
The coefficient $g_{r_0}$ is given as 
\begin{eqnarray}
g_{r_0}&=&
(1/2)\cdot \biggl[
D_{r_0}-\sum_{n=1}^{r_0-1}(-)^n\sum_{r_1=n}^{r_0-1}\sum_{r_2=n-1}^{r_1-1}
\cdots \sum_{r_{n-1}=2}^{r_{n-2}-1}\sum_{r_n=1}^{r_{n-1}-1}
D_{r_0-r_1}D_{r_1-r_2}\cdots \nonumber\\
& &\qquad\qquad\qquad\qquad\qquad\qquad\qquad \times
D_{r_{n-2}-r_{n-1}}D_{r_{n-1}-r_n}D_{r_n}
\biggl] \ , 
\label{A-29}\\ 
D_n&=&(2n)!/(n!)^2 \ . 
\label{A-30}
\end{eqnarray}
The coefficients $g_{r_0}$ from $r_0=1$ to $r_0=10$ are as follows : 
\begin{eqnarray}\label{A-31}
& &g_1=1 \ , \ \quad g_2=1\ ,\ \ \ \quad g_3=2 \ ,\ \ \ \quad g_4=5 \ , 
\ \ \ \ \quad 
g_5=14 \ , \nonumber\\
& &g_6=42 \ , \quad g_7=132\ , \quad g_8=429 \ , \quad g_9=1430 \ , \quad 
g_{10}=4758 \ .
\end{eqnarray}
With the use of the relations (\ref{A-4}), (\ref{A-11}), (\ref{A-12}) 
and (\ref{A-16}) with (\ref{A-17}), we can show the following relation : 
\begin{equation}\label{A-32}
{\hat \tau}_0\ket{t,t_0}=(t_0-t)\ket{t,t_0} \ .
\end{equation}
The proof is easy. We can prove that, for arbitrary $g_r$, the state 
$\ket{t,t_0}$ is an eigenstate of ${\hat \tau}_0$ with the eigenvalue 
$\tau_0$, which is given as 
\begin{eqnarray}\label{A-33}
\tau_0=\sum_{s=1}^{t_0\!-\!t}g_s& &(t_0\!-\!t)!(t_0\!+\!t\!-\!1)!
(2t_0\!-\!2s\!-\!1)!\nonumber\\
& &\times [(t_0\!-\!t\!-\!s)!(t_0\!+\!t\!-\!s\!-\!1)!(2t_0\!-\!2)!]^{-1}
\end{eqnarray}
We intend to determine $g_r$ so as to satisfy the relation 
\begin{equation}\label{A-34}
\tau_0=t_0-t \ .
\end{equation}
The relation (\ref{A-34}) with (\ref{A-33}) can be rewritten as 
\begin{eqnarray}
& &\sum_{s=1}^{r} g_s r!((2t\!-\!1)\!+\!2(r\!-\!s))!((2t\!-\!1)\!+\!r)!
\nonumber\\
& &\qquad\qquad \times 
[(r\!-\!s)!((2t\!-\!1)\!+\!(2r\!-\!1))!((2t\!-\!1)\!+\!(r\!-\!s))!]^{-1}=r \ , 
\label{A-35}\\
& &t_0=t+r \ .
\label{A-36}
\end{eqnarray}
The relation (\ref{A-35}) with $r=1$ gives us 
\begin{equation}\label{A-37}
g_1=1 \ .
\end{equation}
The relation (\ref{A-35}) can be rewritten as 
\begin{eqnarray}\label{A-38}
g_r&=&r((2t\!-\!1)\!+\!2r)![((2t\!-\!1)\!+\!2r)r!((2t\!-\!1)\!+\!r)!]^{-1}
\nonumber\\
& & - \sum_{s=1}^{r-1}g_s((2t\!-\!1)\!+\!2(r\!-\!s))![(r\!-\!s)!((2t\!-\!1)
+(r\!-\!s))!]^{-1}\ .
\end{eqnarray}
With the use of the relation (\ref{A-38}), we can determine $g_r$ 
from $r=2$ to the higher successively. Of course, the relation 
(\ref{A-37}) is used. Then, we can show that $g_r$ does not depend 
on $t$ for any value of $r$. Thus, by putting $t=1/2$ for the 
form (\ref{A-38}), the following recursion formula is derived : 
\begin{equation}\label{A-39}
g_r=D_r - \sum_{s=1}^{r-1} g_s D_{r-s} \ .
\end{equation}
Here, $D_n$ is defined in the relation (\ref{A-30}). The solution of 
the recursion formula (\ref{A-39}) is given in the form (\ref{A-29}). 
The relation (\ref{A-32}) is rewritten as 
\begin{equation}\label{A-40}
{\hat \tau}_0\ket{t,t_0}=({\hat T}_0-{\hat T})\ket{t,t_0} \ .
\end{equation}
Then, we have 
\begin{equation}\label{A-41}
{\hat T}={\hat T}_0-{\hat \tau}_0 \ .
\end{equation}
Of course, ${\hat \tau}_0$ is given in the relation (\ref{A-27}) 
with (\ref{A-28}). It is interesting to see that the operator 
${\hat T}$ can be expressed in terms of the generators ${\hat T}_\pm^0$ 
and ${\hat T}_0$.

\end{document}